%% file: TNNLS-2013-P-0123.tex
\documentclass[onecolumn]{IEEEtran}
\usepackage{calc}
\usepackage[T1]{fontenc}

\usepackage[overload]{empheq}
\usepackage{cleveref}
\usepackage{multirow}
\usepackage{cite}
\usepackage{graphicx,subfigure}
\usepackage{psfrag}
\usepackage{amsmath,amssymb}
\usepackage{color}
\usepackage{tikz}
\usepackage{cleveref}
\usepackage{empheq}
\usepackage{dsfont}
\usepackage{caption}
\usepackage{algorithm}
\usepackage{algpseudocode}
\usepackage[utf8]{inputenc}
\usepackage{setspace}

\include{defs}

\makeatletter
\def\blfootnote{\gdef\@thefnmark{}\@footnotetext}
\def\BState{\State\hskip-\ALG@thistlm}
\makeatother

\begin{document}
\sloppy

%\title{LAGC: Lazy Aggregated Gradient Coding for a Flexible Trade-off Among Communication, Computation, and Wall-clock Time Requirements}
\title{LAGC: Lazily Aggregated Gradient Coding for Straggler-Tolerant and Communication-Efficient Distributed Learning}
\author{Jingjing Zhang and Osvaldo Simeone}  
\maketitle

\thispagestyle{empty}
\begin{abstract}
Gradient-based distributed learning in Parameter Server (PS) computing architectures is subject to random delays due to straggling worker\blfootnote{The authors are with the Department of Informatics, King's College London, London, UK (emails: jingjing.1.zhang@kcl.ac.uk, osvaldo.simeone@kcl.ac.uk).} nodes, as well as to possible communication bottlenecks between PS and workers. Solutions have been recently proposed to separately address these impairments based on the ideas of gradient coding, worker grouping, and adaptive worker selection. This paper provides a unified analysis of these techniques in terms of wall-clock time, communication, and computation complexity measures. Furthermore, in order to combine the benefits of gradient coding and grouping in terms of robustness to stragglers with the communication and computation load gains of adaptive selection, novel strategies, named Lazily Aggregated Gradient Coding (LAGC) and Grouped-LAG (G-LAG), are introduced. Analysis and results show that G-LAG provides the best wall-clock time and communication performance, while maintaining a low computational cost, for two representative distributions of the computing times of the worker nodes. 
%Gradient-based distributed learning in Parameter Server (PS) architectures is subject to random delays due to straggling worker nodes, as well as to possible communication bottleneck between PS and workers. stragglers, heavy communication and computation cost. However, it is difficult to have a unified solution for these problems. For instance, mitigating stragglers and saving computations can be a conflict of interest. In this work, to address these issues, we investigate three different techniques: coding, adaptive selection, and grouping. With a combination of the three techniques, we propose a novel strategy, named as Lazy Aggregated Gradient Coding (LAGC), which explores the trade-off in terms of wall-clock times, communications and computations. We also present a convergence analysis of LAGC and other alternatives. Numerical illustration shows that LAGC provides the best wall-clock time and communication performance, while maintaining a low computation cost, as compared to existing strategies.
\end{abstract}


\section{introduction} \label{intro}

In order to scale machine learning so as to cope with large volumes of input data, distributed implementations of gradient-based methods that leverage the parallelism of first-order optimization techniques are commonly adopted \cite{DCM:12,AAB:15,ZWLA:10}. A standard large-scale distributed computing architecture relies on many parallel worker nodes to perform iterative computations of the gradients and on a central Parameter Server (PS) to aggregate the computed gradients and communicate with the workers \cite{SN:10,MDJ:14}. The PS computing architecture is subject to two key impairments. First, the potentially high tail of the distribution of the computing times at the workers can cause significant slowdowns in wall-clock run-time per iteration due to straggling workers \cite{DB:13}. Second, the communication overhead resulting from intensive two-way communications between the PS and the workers may require significant networking resources to be available in order not to dominate the overall run-time \cite{BH:09}. 

Recently, solutions have been developed that aim at improving robustness to stragglers --- namely Gradient Coding (GC) and grouping \cite{TLDK:17, OFU:19} --- or communication load --- namely adaptive selection \cite{TGSY:18} (see Table~\ref{advan} for a summary). GC, introduced in \cite{TLDK:17}, increases robustness to stragglers by leveraging storage and computation redundancy at the worker nodes as compared to standard (distributed) Gradient Descent (GD) \cite{WBK:16}. With a redundancy factor $r>1$, each worker stores, and computes on, $r$ times more data than with GD. Under GC, given a redundancy factor $r>1$, up to $r-1$ stragglers can be tolerated, while still allowing the PS to exactly compute the gradient at any iteration. GC requires coding the computed gradients prior to communication from the workers to the PS, as well as decoding at the PS. 

\begin{table}[t!]
\caption{Qualitative comparisons with respect to standard (distributed) Gradient Descent (GD)}
\begin{center}
  \begin{tabular}{ | c|| c | c| c| }
    \hline   
                                       & \text{Coding}  & \text{Grouping} & \text{Adaptive selection} \\ [1ex] \hline\hline
    \text{Robustness to stragglers}             & better   &better       & same       \\ \hline
    \text{Communication load}                   & same     & same        & better \\ \hline
	  \text{Computation load}                     & worse    & worse       &better  \\ \hline
  \end{tabular}
\end{center}
\label{advan}
\end{table}

As a special case of GC, given a redundancy factor $r$ equal to the number $M$ of workers, each worker can store the entire dataset. Hence,  the gradient can be obtained from any worker without requiring any coding or decoding operation. In the typical case in which $r$ is smaller than $M$, the same simple procedure can be applied to groups of workers. In particular, given a redundancy factor $r$, the dataset can be partitioned so that each partition is available to all nodes of a group of $r$ workers. The PS can then recover the gradient upon receiving the computations of any server for each group. The outlined grouping scheme can hence tolerate up to $(M/r)(r-1)$ stragglers, which may be significantly larger than $r-1$ when $M\gg r$ \cite{OFU:19}.
%In this paper, we investigate three different techniques: \emph{coding}, \emph{adaptive selection}, and \emph{grouping}. In terms of the aforementioned problems, the quantitative advantages/disadvantages of the three techniques are demonstrated in Table~\ref{advan} for a summary as compared to the standard Gradient Descent  \cite{WBK:16}. In particular, by taking advantage of storage and computation redundancy, coding is capable to offer robustness to stragglers \cite{TLDK:17, ,YA:18}. For instance, with a redundancy factor $r$, up to $r-1$ stragglers can be avoided. Accordingly, the updates from the stragglers to the PS can be ignored, providing some communication reduction. Obviously, this ends up with a heavy computation cost increased by a factor $r$. As compared to coding, grouping is able to tolerate more stragglers, e.g. with $G$ feasible groups, $(r-1)G$ stragglers are tolerable, making the learning process even faster \cite{OFU:19}. This is again with the reliance on redundancy, leading to the same disadvantages as coding. Adaptive selection is a communication-efficient approach by means of selecting the active workers at each iteration \cite{TGSY:18}. Without storage redundancy like GD, it is also computationally favorable but with limited straggler tolerance constrained by the selection operation.

While GC and grouping aim at reducing wall-clock time per iteration by leveraging storage and computation redundancy, the goal of adaptive selection is to reduce the communication and computation loads. This is done by selecting at each iteration a subset of workers to be active \cite{TGSY:18}. Selection is done by predicting at the PS (or in a distributed way at the servers) the servers that are more likely to have informative updates as compared to their latest communicated gradients. The approach, termed Lazily Aggregated Gradient (LAG), is shown in \cite{TGSY:18} to have approximately the same iteration complexity as GD at substantially reduced communication and computation loads. 

In this work, we provide a unified study of coding, grouping, and adaptive selection techniques in terms of wall-clock run-time, communication load, and computation load. Furthermore, in order to combine the benefits of GC and grouping in terms of robustness to stragglers with the communication and computation load gains of adaptive selection of LAG, novel strategies, named Lazily Aggregated Gradient Coding (LAGC) and Grouped-LAG (G-LAG), are introduced. 
%Analysis and results show that G-LAG provides the best wall-clock time and communication performance, while maintaining a low computation cost, as compared to competing strategies.

%In fact, not only stragglers can substantially delay the learning process, the computation redundancy can also slow it down since the run-time per iteration grows with the number of computations. While most existing works studied the number of iterations or epochs, hence, it is important to understand: 1) how different techniques affect the wall-clock time; 2) can we harvest multiple advantages at the same time? In this work, we leverage the three techniques to explore the trade-off with regard to the overheads of wall-clock time, communication, and computation. By combination, we propose a novel strategy that enables a powerful synergy, in the sense that the wall-clock time and communication cost can be significantly reduced, while keeping a relative low computation cost.

\textbf{Related work:}
The original work \cite{TLDK:17} on GC described above has been extended in a number of directions. By coding across the elements of gradient vectors, rather than only across different gradient vectors as in \cite{TLDK:17}, reference \cite{YA:18} proposed a variant of GC that provides a generalized trade-off in terms of straggler tolerance, computation, and communication loads. By using Reed-Solomon codes, reference \cite{WNFB:18} improves the computational complexity of GC. GC techniques that enable the approximate, rather than exact computations of the gradient were studied in \cite{NRDI:18,CPE:17,SOK:19}. References \cite{BWR:19,HZD:19,RAA:19} considered GC for stochastic gradient methods. GC is also part of an active line of research that aims at more generally improving the robustness of distributed computing (see, e.g., \cite{LLPPR:18,LSR:17,YMA:17,FJHDCG:17,SVP:16}). 

Another related line of work addresses the communication overhead of distributed gradient descent. This can be done by using various methods, such as quantization and sparsification \cite{ZLKALZ:17,SYKM:17,WWLZ:18}, which generally entail a performance loss in terms of accuracy; duality-based methods \cite{JSTTKHJ:14,CJMVMPM:17} and approximate Newton-type methods \cite{SSZ:14,ZL:15}, both of which increase the computational load in order to guarantee a given accuracy; and adaptive batch size selection \cite{MCADAM:18}, which can reduce iteration complexity and communication load at a larger computation cost per iteration than conventional gradient methods.

 %\textcolor[rgb]{1,0,0}{In contrast, references \cite{TGSY:18} can reduce both communication and computational loads by operating on an increasing subset of samples at each iteration and by adaptive selection, respectively.}

%In contrast, the gradient-based LAG scheme introduced in \cite{TGSY:18} can reduce both communication and computational loads via adaptive selection.

Finally, computation complexity in large-scale machine learning applications has also attracted extensive attention. The standard approach is Stochastic Gradient Descent (SGD), which trains using a random mini-batch of samples at each iteration \cite{B:10}. A large number of variants of SGD have been proposed that present different trade-offs among accuracy, communication load, and computation complexity \cite{LO:07}. Typical solutions include quantization and sparsification \cite{DDJRM:17,BWKA:18}. Of particular practical relevance are also asynchronous schemes \cite{JDM:86}, whereby updates by different workers can be aggregated as they are produced with limited need for locking mechanism \cite{DGGDN:17,RRWF:11}. Alternatives to SGD are explored in \cite{MG:15,JRJ:17}.
%lock-less asynchronous SGC \cite{RRWF:11}. 
%By taking advantage of sparsity in data, reference \cite{RRWF:11} shows that SGD can be implemented without any locking, and hence some associated overhead can be eliminated. Using approximation, comparing with SGD, a much faster approach called K-FAC is proposed in \cite{JRJ:17}. Furthermore, to exploit parallel computing resources, asynchronous distributed computing algorithms on SGD are also investigated \cite{JDM:86}. }

%In page 2, the part related to Related work needs to be modified. It is suggested to write this part in order to have stronger literature review that the reader would be familiar with some of the studies related to this manuscript. For the distributed first order methods, the authors can also talk about the studies [1, 2, 3]. In line 20, the authors mentioned 1 about two distributed Newton-type methods, however, there are many other studies such as [4, 5, 6] which are efficient in terms of computation and communication.

\textbf{Main contributions:}
This paper studies gradient-based optimization in a distributed PS architecture that enables GC, grouping, and adaptive selection. An analysis of wall-clock run-time complexity, communication complexity, and computation complexity is provided to measure the performance of GC, LAG, and of the newly proposed LAGC and G-LAG methods. The main contributions are summarized as follows.

1) A novel strategy, named LAGC, is proposed that is able to leverage the advantages summarized in Table~\ref{advan} of GC and LAG via a specific integration of the two techniques. As a special case, we also consider a scheme that only uses grouping and adaptive selection, hence not requiring coding, which is referred to G-LAG; 

2) We provide an analysis of time complexity, communication complexity, and computation complexity for the discussed strategies, namely GD, GC, and G-LAG under the standard assumption of smooth convex loss. We specifically illustrate the trade-off among these metrics under Pareto and exponential distributions for the random computing times of the workers. These distributions are representative of high and low tails, respectively, for the computing times;

3) Finally, we present numerical results on baseline regression tasks so as to study the evolution of the accuracy, as well as communication and computation loads, as functions of wall-clock time.

The rest of the paper is organized as follows. Section \ref{sec:model} describes the PS architecture and the adopted performance metrics. Under this framework, GC and LAG are reviewed in Section \ref{background}, and the proposed strategies are presented in Section \ref{proposed}. Section \ref{analysis} provides the analysis of the adopted metrics along with some numerical illustrations. Section \ref{numerical} presents some numerical examples for a regression task. Finally, Section \ref{conclusion} concludes the work and also highlights future research directions. 

\section{System Model} \label{sec:model}

\begin{figure}[t!] 
  \centering
\includegraphics[width=0.35\columnwidth]{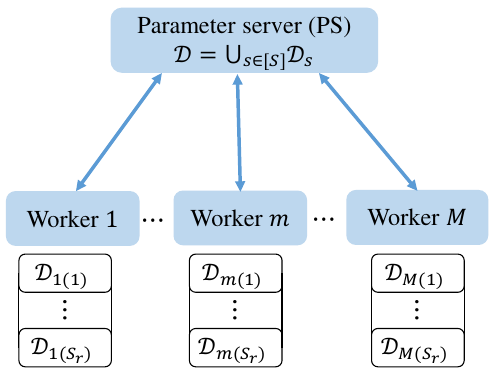}
\caption{Parameter Server (PS) model with storage redundancy $r$.}
\label{fig:model}
\end{figure}

We are given a training dataset $\mathcal{D}=\{\zv_n=(\xv_n,y_n)\}_{n=1}^{N}$, where the explanatory vector $\xv_n\in \R^{d}$ contains $d$ covariates and the label $y_{n}\in \F$ takes values in a finite discrete set $\F$. The objective is to learn a model parameter vector $\pmb{\theta}\in \R^{p}$ by minimizing the training loss
\begin{align} \label{loss}
L(\mathcal{D}; \pmb{\theta})=\mathop{\sum}_{\zv_n\in\mathcal{D}} \ell(\zv_n;\pmb{\theta}),
\end{align}
where $\ell(\zv_n;\pmb{\theta})$ is a loss function that depends on the hypothesis class and on the performance criterion of interest, e.g., quadratic error or cross-entropy. To tackle the minimization of function $L(\mathcal{D}; \pmb{\theta})$ over vector $\pmb{\theta}$, we consider methods based on approximate Gradient Descent (GD) steps, whereby parameter $\pmb{\theta}$ is updated iteratively by following the rule
\begin{align} \label{update}
\pmb{\theta}^{i+1}=\pmb{\theta}^{i}-\alpha \hat{\gv}(\pmb{\theta}^i),
%=\pmb{\theta}^{i}-\alpha \big(\gv_{\mathcal{D}^i}(\pmb{\theta}^{i})+\gv_{\bar{\mathcal{D}}^i}(\{\pmb{\theta}^{i-k}\}_{k=1}^{i-1})\big),
\end{align}
with $\alpha$ being the stepsize; superscript $i$ indicating the iteration index; and $\hat{\gv}(\pmb{\theta}^i)$ being an estimate of the exact gradient $\gv(\pmb{\theta}^i)=\nabla L(\mathcal{D};\pmb{\theta}^i)$. Note that we focus on full gradient techniques that aim at linear convergence rates in terms of number of iterations, and we do not consider stochastic GD methods, which instead can only achieve sub-linear convergence rates (see, e.g., \cite{WBK:16}). 
%the evaluated gradient over all dataset $\mathcal{D}$ by combining the computed results from the workers at current iteration $i$ and the outdated gradient $\hat{\gv}(\pmb{\theta}^{i-1})$ from the previous iteration $i-1$.

%$\mathcal{D}^i\subseteq \mathcal{D}$ being a subset, or minibatch, of training examples; 
%\begin{align} \label{gradient}
%\gv_{\mathcal{D}^i}(\pmb{\theta}^i)=\mathop{\sum}_{\zv_n\in\mathcal{D}^i} \nabla \ell(\zv_n;\pmb{\theta}^i)
%\end{align}
 %being the gradient of the loss function evaluated over a subset $\mathcal{D}^i\in\mathcal{D}$ of training data; and 
%\begin{align} \label{oldgradient}
%\gv_{\bar{\mathcal{D}}^i}(\{\pmb{\theta}^{i-1}\}_{k=1}^{i-1})=\mathop{\sum}_{\zv_n\in\bar{\mathcal{D}}^i} \nabla \ell(\zv_n;\{\pmb{\theta}^{i-1}\}_{k=1}^{i-1})
%\end{align}
 %being the outdated gradient as a function of the old versioned model parameters $\{\pmb{\theta}^{i-k}\}_{k=1}^{i-1}$ over the subset $\bar{\mathcal{D}}^i=\mathcal{D}\backslash\mathcal{D}^i$ of the rest training data. When $\bar{\mathcal{D}}^i=\emptyset$ and hence $\gv_{\bar{\mathcal{D}}^i}\big(\{\pmb{\theta}^{i-1}\}_{k=1}^{i-1}\big)=\mathbf{0}$, we refer to $\gv(\pmb{\theta}^i)=\gv_{\mathcal{D}}(\pmb{\theta}^i)$ over the entire dataset $\mathcal{D}$ as the full gradient.
 %write $\gv(\pmb{\theta}^i)=\gv_{\mathcal{D}}(\pmb{\theta}^i)$ for the full gradient over the entire dataset $\mathcal{D}$, i.e., with $\bar{\mathcal{D}}^i=\emptyset$ and hence $\gv_{\bar{\mathcal{D}}^i}\big(\{\pmb{\theta}^{i-1}\}_{k=1}^{i-1}\big)=\mathbf{0}$.

A Parameter Server (PS) framework is commonly adopted to run Gradient Descent (GD) using parallel workers. As illustrated in Fig.~\ref{fig:model}, the PS has access to the entire dataset $\mathcal{D}$ and it can communicate with $M$ workers, which are denoted by set $\mathcal{M}=\{1,2,\dots,M\}$. Note that the workers are part of the same computation system, and hence communication is not subject to privacy constraints. In the following, we describe a framework for a PS-based implementation of updates \eqref{update} that allows us to study in a unified way GC \cite{TLDK:17} and (an equivalent variant of) LAG \cite{TGSY:18}, and to generalize both via the proposed LAGC strategy. 

Prior to the start of training, the dataset $\mathcal{D}$ is partitioned into $S$ subsets $\mathcal{D}_1,\dots,\mathcal{D}_S$ of equal size, with the partial gradient of each data partition $\mathcal{D}_s$ defined as 
\begin{align}\label{gradient}
\gv_{s}(\pmb{\theta}^i)=\mathop{\sum}_{\zv_n\in\mathcal{D}_s} \nabla \ell(\zv_n;\pmb{\theta}^i).
\end{align}
As seen in Fig.~\ref{fig:model}, to parallelize the computation of the gradient in \eqref{update}, each worker $m\in\mathcal{M}$ is assigned $S_r=rS/M$ partitions for some integer $1\leq r\leq M$. The partitions assigned to worker $m$ are denoted as $\mathcal{D}_{m(1)},\dots,\mathcal{D}_{m(S_r)}$, where $m(j)\in\{1,\dots,S_r\}$ for $j=1,\dots,S_r$. Integer $r$ is referred to as the \emph{storage redundancy}, since, with an even arrangement of the data partitions, each partition is replicated $r$ times across workers. Note that the choice $r=M$ implies that the entire dataset $\mathcal{D}$ can be stored at each worker. 

\begin{figure}[t!] 
  \centering
\includegraphics[width=0.42\columnwidth]{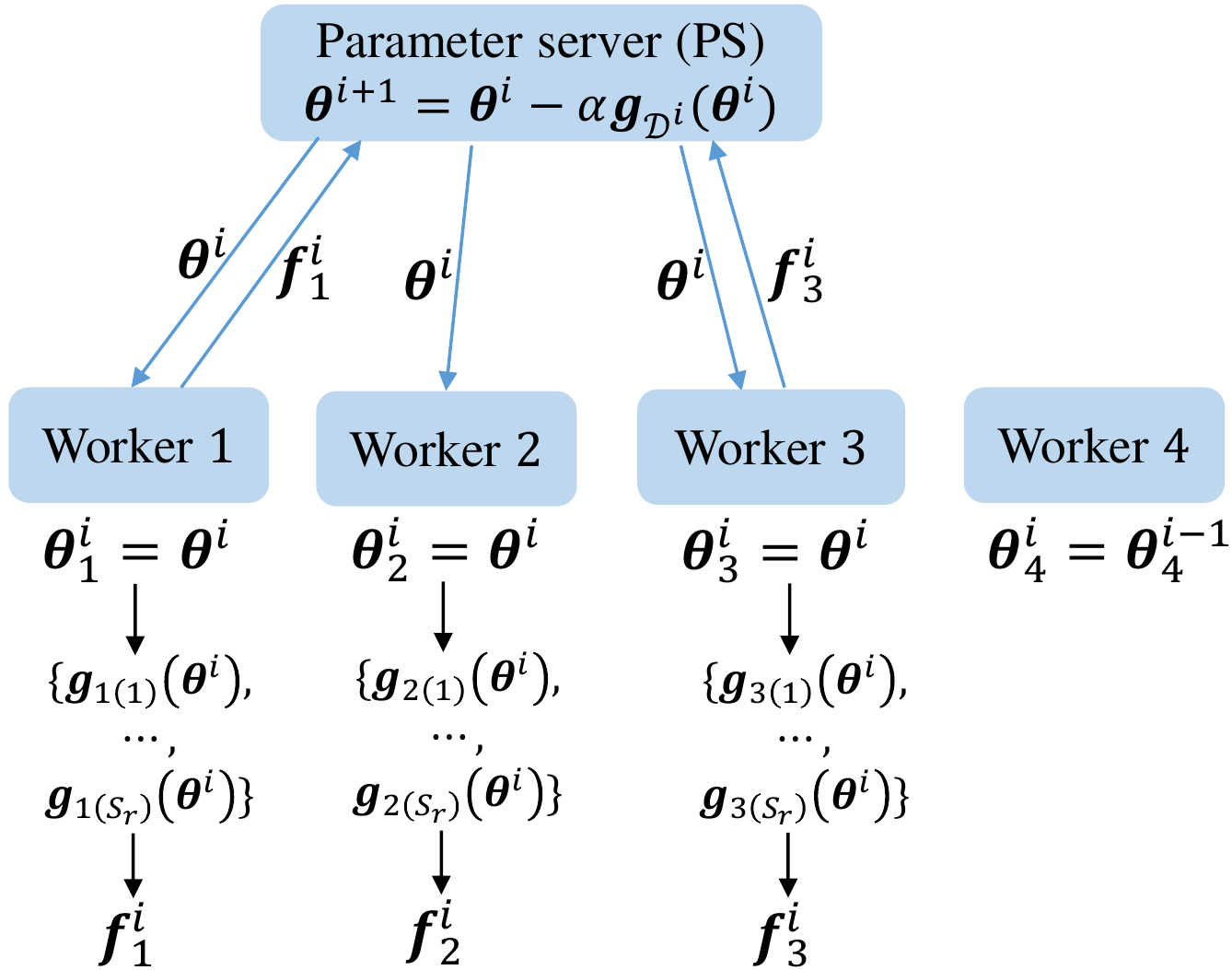}
\caption{Illustration of the training protocol with $M=4$ workers, $\mathcal{M}_D^i=\{1,2,3\}$ and $\mathcal{M}_U^i=\{1,3\}$ at the $i$th iteration.}
\label{fig:example}
\end{figure}

To elaborate on the communication and computation protocol, we define $\pmb{\theta}^{i}_m$ as the version of the model parameter that worker $m\in\mathcal{M}$ has available at iteration $i$ prior to computation. At each iteration $i$, a subset of workers, denoted by $\mathcal{M}_D^i\subseteq\mathcal{M}$, is selected to download the model parameter $\pmb{\theta}^i$ from the PS. As illustrated in Fig.~\ref{fig:example}, each selected worker $m\in\mathcal{M}_D^i$ sets $\pmb{\theta}^{i}_m=\pmb{\theta}^{i}$, and computes the local partial gradients $\gv_{m(1)}(\pmb{\theta}^{i}),\dots,\gv_{m(S_r)}(\pmb{\theta}^{i})$ over all assigned data partitions. In contrast, each non-selected worker $m\in\mathcal{M}\backslash\mathcal{M}_D^i$ does not download the current model parameter from the PS, and it reuses the locally available parameter vector $\pmb{\theta}_m^{i-1}$ from the previous iteration by setting $\pmb{\theta}^{i}_m=\pmb{\theta}^{i-1}_m$. As a result, the local parameter vector $\pmb{\theta}^{i}_m$ is different from global parameter vector $\pmb{\theta}^{i}$ for all workers $m\in\mathcal{M}\backslash\mathcal{M}_D^i$.

%In contrast, each non-selected worker $m\in\mathcal{M}\backslash\mathcal{M}_D^i$ sets $\pmb{\theta}^{i}_m=\pmb{\theta}^{i-1}_m$. Note that $\pmb{\theta}^{i}_m$ is different from $\pmb{\theta}^{i}$ for $m\in\mathcal{M}\backslash\mathcal{M}_D^i$.

The wall-clock time $T_m^i$ required to complete the computation of all local gradients at each worker $m\in\mathcal{M}_D^i$ is random, with mean $\eta r$, for some $\eta>0$, proportional to the worker load, which is in turn proportional to the storage redundancy $r$. Variables $\{T_m^i\}_{m\in\mathcal{M}_D^i}$ are assumed to be i.i.d. across the worker index $m$ and iteration index $i$. This standard assumption is motivated by the fact that, in most deployments, the workers are similar machines that operate independently (see, e.g., \cite{MDJ:14,TLDK:17,TGSY:18,YA:18}).
In this paper, we will consider as representative examples an exponential distribution with mean $\eta r$ and a Pareto distribution with scale-shape parameter pair $(\eta r(\beta-1)/\beta,\beta)$, where $\eta>0$ and $\beta>1$ are constants. The latter has a higher tail than the former, implying a larger probability of straggling workers. 

In order to reduce the time per iteration, as illustrated in Fig.~\ref{fig:example}, the PS may only wait for a subset $\mathcal{M}_U^i\subseteq\mathcal{M}_D^i$ of active workers to complete their computations. Each worker $m\in\mathcal{M}_U^i$ uploads a function $\fv^i_m\in \R^p$ of the computed partial gradients $\{\gv_{m(j)}(\pmb{\theta}^{i})\}_{j=1}^{S_r}$ at current iteration back to the PS. Based on the received functions $\{\fv^i_m\}_{m\in\mathcal{M}_U^i}$, and possibly also previously received functions $\{\fv^{j}_m\}_{m\in\mathcal{M}_U^{j}}$, with $ j<i$, the PS computes an estimate $\hat{\gv}(\pmb{\theta}^{i-1})$ of the gradient. Note that each worker $m$ in subset $\mathcal{M}_D^i\backslash\mathcal{M}_U^i$ downloads the model parameter $\pmb{\theta}^{i}$ but it does not upload a local update due to the excessive time elapsed for its computation. Similar to each worker in subset $\mathcal{M}\backslash\mathcal{M}_D^i$, the workers in subset $\mathcal{M}_D^i\backslash\mathcal{M}_U^i$ also reuse the outdated gradients $\{\mathbf{g}_{m(j)}(\pmb{\theta}^{i-1})\}_{j=1}^{S_r}$ for the local parameter vector $\pmb{\theta}^{i}_m=\pmb{\theta}^{i-1}_m$.

%It can be observed that the training dataset, denoted by $\mathcal{D}^i$, used for computation at current iteration may be partial, given as $\mathcal{D}^i=\bigcup_{m\in\mathcal{M}_U^i}\bigcup_{i\in[S_r]}\mathcal{D}_{m(i)}$. The functions are designed so that, by using the received vectors $\{\fv_m^i\}_{m\in\mathcal{M}_U^i}$ and the outdated global gradient $\hat{\gv}(\pmb{\theta}^{i-1})$, the PS can recover gradient $\hat{\gv}(\pmb{\theta}^{i})$. 
The PS then updates the model parameter $\pmb{\theta}^i$ via the rule in \eqref{update}. The PS also keeps track of the workers' parameter vectors $\{\pmb{\theta}^{i}_m\}_{m\in\mathcal{M}}$. The next iteration $i+1$ then starts with the workers in subset $\mathcal{M}_D^{i+1}$ downloading the updated model $\pmb{\theta}^{i+1}$ from the PS. The training continues until a convergence criterion is satisfied or a maximum number of iterations is reached. 

%As we will discuss, the determination of the two subsets $\mathcal{M}_D^i$ and $\mathcal{M}_U^i\subseteq\mathcal{M}_D^i$ at each iteration is carried out based on different criteria for LAG \cite{TGSY:18}, GC\cite{TLDK:17}, and the scheme proposed here.

Throughout this paper, by following \cite{TGSY:18,DGGDN:17,BEJ:16}, we make the following standard assumptions on the local loss functions $L_{s}({\pmb{\theta}})=\mathop{\sum}_{\zv_n\in\mathcal{D}_s} \ell(\zv_n;\pmb{\theta})$ for each data partition $\mathcal{D}_s$ and on the overall loss function $L(\pmb{\theta})=\sum_{s=1}^{S}L_s(\pmb{\theta})=L(\mathcal{D}; \pmb{\theta})$.\\
\emph{Assumption 1:} The local loss functions $L_{s}({\pmb{\theta}})$ are $L_s$-smooth for $s=1,\dots,S$, for some $L_{s}>0$, that is, we have the inequalities 
\begin{align} \label{smooth}
||\nabla L_s(\pmb{\theta})-\nabla L_s(\pmb{\theta}')||^2\leq L_s^2||\pmb{\theta}-\pmb{\theta}'||^2
\end{align}
for all $\pmb{\theta}$ and $\pmb{\theta}'\in \R^p$; and for $s=1,\dots,S$, and the loss function $L(\pmb{\theta})$ is $L$-smooth with $L\leq \sum_{s=1}^{S}L_s$.  From \eqref{smooth}, the smoothness parameters $L_s$ determine the rate of variability of the gradient of each local loss function $L_{s}({\pmb{\theta}})$. The parameters $\{L_s\}$ are assumed to be known, and they can be practically estimated using techniques such as those discussed in \cite{WZ:96}. \\ 
\emph{Assumption 2:} $L(\pmb{\theta})$ is $\mu$-strongly convex, or more generally, it satisfies the Polyak \L{}ojasiewicz (PL) condition for some $\mu>0$, i.e., it satisfies the inequality 
\begin{align} \label{assum:2}
2\mu(L(\pmb{\theta})-L(\pmb{\theta}^*)) \leq ||\gv({\pmb{\theta}})||^2
\end{align}
for all ${\pmb{\theta}}\in\R^p$, where $\pmb{\theta}^*$ is a minimum of $L(\pmb{\theta})$. Note that the minimum $\pmb{\theta}^*$ is unique for strongly convex function. From \eqref{assum:2}, the norm of the gradient can be used as a measure of distance to the optimal value of the loss function.
 
\subsection{Performance Metrics}

We are interested in studying the performance in terms of training accuracy, communication load, and computation load as a function of the wall-clock time. To this end, we start by defining the number $I$ of iterations of the SGD rule \eqref{update} carried out by time $t$ as 
\begin{align} \label{comple:iter}
I(t)=\max\bigg\{I:\sum_{i=1}^{I}\max_{m\in\mathcal{M}_U^i}\{T_m^i\}\leq t\bigg\}.
\end{align}
Note that $I(t)$ is a random variable due to the randomness of the times $T_m^i$ and of the subsets $\mathcal{M}_U^i$. Furthermore, the quantity $\max_{m\in\mathcal{M}_U^i}\{T_m^i\}$ represents the time per iteration since the PS waits for all the workers in subset $\mathcal{M}_U^i$ to finish their respective computations. For each iteration $i$, the \emph{communication load} is defined as the sum of the numbers $|\mathcal{M}_D^i|$ of workers that download model parameters from the PS and of the numbers $|\mathcal{M}_U^i|$ of workers that upload the computed gradients to the PS. It follows that the \emph{communication load} $C(t)$ as a function of time $t$ is given as 
 \begin{align} \label{load:comm}
C(t)=\sum_{i=1}^{I(t)}\big(|\mathcal{M}_D^i|+|\mathcal{M}_U^i| \big).
\end{align}
%where $\lambda$ is the ratio between the size of the local gradient each active worker uploads and the size of the global gradient $\hat{\gv}(\pmb{\theta}^i)$. 
We also define the \emph{computation load} as the total number of gradients per data point computed at the workers. The computation load at time $t$ is given by
 \begin{align} \label{load:comp}
P(t)=\sum_{i=1}^{I(t)}\frac{r}{M}|\mathcal{M}_D^i|,
\end{align}
since each worker in set $\mathcal{M}_D^i$ computes $rN/M$ local gradients. Note that the workers in $\mathcal{M}_U^i\backslash\mathcal{M}_D^i$ may compute only partially the local gradients, but here we do not make this distinction since, in practice, this would require additional signaling from PS to the workers. Finally, the \emph{training loss optimality gap} at time $t$ is given by 
\begin{align} \label{def:loss}
L(t)=L(\pmb{\theta}^{I(t)})-L(\pmb{\theta}^*).
\end{align} 

Beside the random tuple $\big(L(t),C(t),P(t)\big)$, we also consider the \emph{time complexity}, the \emph{communication complexity}, and the \emph{computation complexity}, which are summary metrics that measure as the average time, communication load, and computation load needed to ensure an optimality gap equal to $\epsilon>0$. Accordingly, defining as the (random) number of iterations needed to obtain an $\epsilon$-optimality gap, also known as iteration complexity \cite{TGSY:18}, $I_\epsilon=\min\{I:||L(\pmb{\theta}^{I})-L(\pmb{\theta}^*)||^2\leq \epsilon\}$, the wall clock time complexity is defined as 
\begin{align}\label{def:time}
\bar{T}_{\epsilon}=\mathrm{E}\bigg[\sum_{i=1}^{I_\epsilon}\max_{m\in\mathcal{M}_U^i}\{T_m^i\}\bigg],
\end{align} 
the communication complexity as
\begin{align}\label{def:commu}
\bar{C}_{\epsilon}=\mathrm{E}\bigg[\sum_{i=1}^{I_\epsilon}|\mathcal{M}_D^i|+|\mathcal{M}_U^i|\bigg],
\end{align}
and the computation complexity as
\begin{align} \label{def:compu}
\bar{P}_{\epsilon}=\mathrm{E}\bigg[\sum_{i=1}^{I_\epsilon}\frac{r}{M}|\mathcal{M}_D^i|\bigg].
\end{align} 
We note that we have included in the wall-clock time only the durations of the computation steps, hence excluding the contribution of communications. This allows to more clearly highlight the trade-off between computing and communication. A compound wall-clock run-time metric that accounts for both computation and communication can be easily derived from the results in this paper. 

\section{Background} \label{background}
 
In this section, we review the state-of-art techniques GC \cite{TLDK:17} and LAG \cite{TGSY:18}.

\subsection{Gradient Coding (GC)} \label{sec:GC}

\begin{figure}[t!] 
  \centering
\includegraphics[width=0.32\columnwidth]{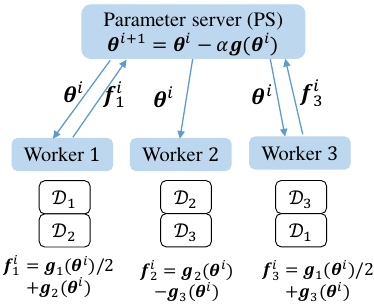}
\caption{Gradient Coding (GC): at current iteration $i$, the PS can recover the full gradient $\gv(\pmb{\theta}^{i})$ by  aggregating the functions $\fv_1^i$ and $\fv_3^i$ received from the fastest workers 1 and 3.}
\label{fig:cg}
\end{figure}

GC, introduced in \cite{TLDK:17}, is an exact full-gradient descent approach, implementing rule \eqref{update} with $\hat{\gv}(\pmb{\theta}^{i})=\gv(\pmb{\theta}^{i})$, that aims at mitigating straggling workers by leveraging storage and computational redundancy. 
%In terms of the metrics defined above, GC can reduce wall clock time and communication complexities at the cost of a larger computation complexity. 
Prior to training, GC replicates each data partition $r>1$ times across the workers. To this end, dataset $\mathcal{D}$ is first divided into $S=M$ partitions $\{\mathcal{D}_s\}_{s=1}^{S}$. Each partition $\mathcal{D}_s$ is stored at $r$ workers, and we have $S_r=r$ partitions at each worker. Specifically, worker $m$ stores partition $\mathcal{D}_{[m+i]_M}$ with $i=0,\dots,r-1$, with $[m]_M=\bmod_M(m-1)+1$ and $\bmod_M(\cdot)$ being the modulo-$M$ operation (see Fig.~\ref{fig:cg} for an example).
%Then the $S$ partitions are assigned to the $M$ workers of any order as $\mathcal{D}_{1}, \cdots,\mathcal{D}_{M}$. The assignment repeats this operation in a cyclic manner for another $r-1$ times, i.e., for each $i=2,\cdots,r$, we have $\mathcal{D}_{i}, \cdots, \mathcal{D}_{M},\cdots, \mathcal{D}_{i-1}$ for the $M$ workers of the same order.
At each iteration $i$, all the workers download the model $\pmb{\theta}^{i}$ from the PS to execute the computations, i.e., we have $\mathcal{M}_D^i=\mathcal{M}$ (recall Fig.~\ref{fig:example}). Note that, for GC, we hence have the local parameter given as $\pmb{\theta}^{i}_m=\pmb{\theta}^{i}$ for all the workers. The PS waits only the fastest $F$ workers to finish their computations, yielding the subset $\mathcal{M}_U^i=\{m\in \mathcal{M}:T^i_m\leq T^i_{F:M}\}$, where $T^i_{F:M}$ is the $F$th order statistic of the variables $\{T_{m}^i\}_{m=1}^{M}$. To enable the recovery of the gradient $\gv(\pmb{\theta}^{i})$ at the PS, each of the worker in $\mathcal{M}_U^i$ sends a designed linear combination $\fv_m^i$ of the computed partial gradients $\{\gv_{m(i)}(\pmb{\theta}^{i})\}_{i=1}^{S_r}$ to the PS. The PS then computes a linear combination of the vectors $\{\fv_m^i\}_{m\in\mathcal{M}_U^i}$ so as to recover the full gradient $\gv(\pmb{\theta}^{i})$. In order to guarantee the existence of linear encoding and decoding functions that enable the recovery of the full gradient $\gv(\pmb{\theta}^{i})$ for any set of $F$ workers, the inequality $F\geq M-r+1$ is necessary and sufficient \cite{TLDK:17}. 

As a special case of GC, if $r=M$, the entire dataset can be stored at each worker. In this case, the PS can wait for the fastest worker only ($F=1$), and no coding and decoding operations are needed.

\emph{Example:} Consider $M=3$ workers, $S_r=2$ data partitions at each worker, and storage redundancy $r=2$. GC allows the PS to wait only for the $F=2$ fastest workers since $F=2\geq M-r+1$. As shown in Fig.~\ref{fig:cg}, this is done by splitting dataset $\mathcal{D}$ into $S=M=3$ partitions $\mathcal{D}_1,\mathcal{D}_2$, and $\mathcal{D}_3$. Worker $m=1,2$ stores $S_r=r=2$ partitions $\mathcal{D}_{m}$ and $\mathcal{D}_{m+1}$, while and worker 3 stores $\mathcal{D}_{3}$ and $\mathcal{D}_{1}$. 
%At each iteration $i$, all the three workers download the parameter $\pmb{\theta}^{i}$. The PS waits the fastest two workers, e.g., $\mathcal{M}_U^i=\{m:T^i_m\leq T^i_{2:3}\}=\{1,3\}$, to finish their computations. With the assigned data partitions, worker 1 and 3 compute $\gv_1(\pmb{\theta}^{i})$, $\gv_2(\pmb{\theta}^{i})$ and $\gv_3(\pmb{\theta}^{i})$, $\gv_1(\pmb{\theta}^{i})$, respectively. 
The three workers compute the linear functions indicated in Fig.~\ref{fig:cg}. It can be easily seen that, by summing functions recevied from any two workers, the PS can recover the exact full gradient $\gv(\pmb{\theta}^{i})$. 
%up $\fv_1^{i}=\gv_1(\pmb{\theta}^{i})/2+\gv_2(\pmb{\theta}^{i})$ from worker 1 and $\fv_3^{i}=\gv_1(\pmb{\theta}^{i})/2+\gv_3(\pmb{\theta}^{i})$ from worker 3, the PS recovers the exact full gradient $\gv(\pmb{\theta}^{i})=\sum_{m\in\mathcal{M}_U^i}\fv_m^{i}=\fv_1^{i}+\fv_3^{i}=\sum_{m\in[3]}\gv_m(\pmb{\theta}^{i})$. The PS then computes $\pmb{\theta}^{i+1}$ and broadcasts it to all the three workers to move on to the next iteration.  

\subsection{Lazily Aggregated Gradient (LAG)} 

Lazily Aggregated Gradient (LAG), proposed in \cite{TGSY:18}, is an approximate gradient descent scheme that judiciously selects the subset $\mathcal{M}_D^i$ of active workers at each iteration in order to reduce communication and computation loads. Unlike GC, LAG does not require storage redundancy, and hence we have $r=1$. We can also set without loss of generality $S=M$ and assign each worker $m$ a disjoint data partition $\mathcal{D}_m$, so that $S_r=1$. The PS determines the subset $\mathcal{M}_D^i$ of active workers at each iteration $i$ on the basis of the estimated change in the gradient for the local loss function $L_m(\pmb{\theta})$ corresponding to the data partition at worker $m$ as compared to the latest communicated gradient from the worker. Note that, since the subset $\mathcal{M}_D^i$ is determined irrespective of the realization of the computation times, unlike GC, LAG is not tolerant to stragglers. 
 %However, like GC, LAG can reduce the communication load by requiring only the selected workers to communicate with the PS. Unlike GC, this communication load reduction does not call for any computation redundancy. Furthermore, it comes at no cost in term of convergence rate with respect to full-gradient descent.

In the following, we provide a more detailed description of LAG. We specifically follow the LAG-PS strategy introduced in \cite{TGSY:18}. More precisely, in order to highlight the common elements with GC, the scheme described here is functionally equivalent to LAG-PS, but it differs from it in terms of the way operations are split between encoding functions computed at the workers and decoding functions evaluated at the PS (see Remark \ref{remark:LAG} below for details). 

%To better fit in the PS framework under study and to generalize both LAG and GC, the algorithm sketched below is different from the original LAG in \cite{TGSY:18} but solely in the information the PS receives and stores. Hence, they are still equivalent in terms of convergence rate, and the defined performance metrics, as discussed below. In particular, the dataset $\mathcal{D}$ is divided into $S=M$ partitions $\mathcal{D}_1,\cdots,\mathcal{D}_M$, and each worker $m$ takes a distinct partition $\mathcal{D}_{m}$. 
At each iteration $i$, the PS first determines the subset $\mathcal{M}_D^i$ by checking the following condition for each worker $m$
\begin{align} \label{con:lazy}
L_m^2\Big|\Big|\pmb{\theta}^{i-1}_m-\pmb{\theta}^i\Big|\Big|^2 \geq \frac{\xi}{\alpha^2 M^2 D} \sum_{d=1}^{D}\Big|\Big|\pmb{\theta}^{i+1-d}-\pmb{\theta}^{i-d}\Big|\Big|^2,
\end{align}
where we recall that $L_{m}$ is the smoothness constant of the local function $L_m(\pmb{\theta})$, while $\xi<1$ is some constant. By \eqref{smooth}, the left-hand side of \eqref{con:lazy} represents a bound on the change in the gradient squared norm expected for the local loss at worker $m$ as compared to the last available gradient from the worker. The right-hand side of \eqref{con:lazy}, by the rule \eqref{update}, represents the per-server average contribution over the most recent $D$ iterations to the approximate gradient norm squared $(1/D) \sum_{d=1}^{D}||\hat{\gv}(\pmb{\theta}^{i-d})||^2$, scaled by a parameter $\xi$. The PS selects the workers that satisfy condition \eqref{con:lazy}, i.e., the workers that are expected to have a sizeable difference between the current local gradient and their more recently computed gradients, and hence may contribute more significantly to the model update \eqref{update}. This yields the subset $\mathcal{M}_D^i=\{m\in\mathcal{M}: \eqref{con:lazy}~\text{holds}\}$.

 %the PS selects the workers in set $\mathcal{M}^t_D$ to execute computation. To this end, the PS stores a copy $\pmb{\theta}^{t}_m$ of current iteration worker $m\in\mathcal{M}$ knows about parameter $\pmb{\theta}$, and recent updates $\{\pmb{\theta}^{t-d}\}_{d=0}^{D}$, with some integer $D$. With the stored information, the PS has a condition check for each worker $m\in\mathcal{M}$, which is given as
%\begin{align} \label{con:lazy}
%L_m^2\Big|\Big|\pmb{\theta}^{i-1}_m-\pmb{\theta}^t\Big|\Big|^2 \leq \frac{1}{\alpha^2 M^2} \sum_{d=1}^{D}\xi_d \Big|\Big|\pmb{\theta}^{i+1-d}-\pmb{\theta}^{i-d}\Big|\Big|,
%\end{align}
%where $\pmb{\theta}^{t-1}_m$ is the stored copy that worker $m$ knows about $\pmb{\theta}$ at iteration $t-1$ by the PS, 

With the received parameter $\pmb{\theta}^i$, each worker $m\in\mathcal{M}_D^{i}$ computes the local gradient $\gv_m(\pmb{\theta}^{i})$ in \eqref{gradient} and then sends the result $\fv^i_m=g_m(\pmb{\theta}^i)$ to the PS. In contrast, for each unselected user in subset $\mathcal{M}\backslash \mathcal{M}_D^i$, which does not compute an update, the outdated partial gradients $\{\mathbf{g}_m(\pmb{\theta}^{i-1}_m)\}_{m\in\mathcal{M}\backslash\mathcal{M}_D^{i}}$ from iteration $i-1$ is reused. Note that we have $\mathcal{M}_U^{i}=\mathcal{M}_D^{i}$. By combining both the computed gradients and the outdated gradients, the PS estimates the full gradient
\begin{align}\label{update:LAG}
\hat{\gv}(\pmb{\theta}^{i})=\sum_{m\in{\mathcal{M}^{i}_D}} \gv_m(\pmb{\theta}^{i})+\sum_{m\in{\mathcal{M}\backslash \mathcal{M}^{i}_D}}\gv_m(\pmb{\theta}^{i-1}_m).
\end{align}
The PS computes $\pmb{\theta}^{i+1}$, and updates the variables $\{\pmb{\theta}_m^i=\pmb{\theta}^{i}\}_{m\in\mathcal{M}_D^{i}}$ and $\{\pmb{\theta}_m^i=\pmb{\theta}_m^{i-1}\}_{m\in{\mathcal{M}\backslash \mathcal{M}_D^{i}}}$ before moving to the next iteration.

\begin{figure}[t!] 
  \centering
\includegraphics[width=0.28\columnwidth]{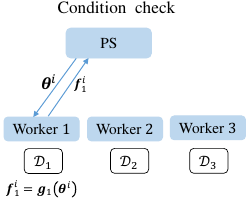}
 \caption{LAG: At current iteration $i$, only worker 1 satisfies Condition \eqref{con:lazy}, and hence only partial gradient $\fv_1^i=\gv^i_1$ is communicated to the PS for update.}
\label{fig:lazy}
\end{figure}

\emph{Example:} Consider again $M=3$ workers. As illustrated in Fig.~\ref{fig:lazy}, at each iteration $i$, the PS checks condition \eqref{con:lazy} for each worker. Assume that only worker 1 satisfies it, and hence we have $\mathcal{M}_D^i=\mathcal{M}_U^i=\{1\}$. Therefore, only worker 1 downloads the current model $\pmb{\theta}^i$ from the PS. It then computes the local gradient $\gv_1(\pmb{\theta}^{i})$ and uploads $\fv_1^i=\gv_1(\pmb{\theta}^{i})$ to the PS. Combining with the outdated partial gradients $\gv_2(\pmb{\theta}^{i-1}_2)$ and $\gv_3(\pmb{\theta}^{i-1}_3)$ of worker 2 and 3, the PS recovers the estimate $\hat{\gv}(\pmb{\theta}^{i})=\gv_1(\pmb{\theta}^{i})+\gv_2(\pmb{\theta}^{i-1}_2)+\gv_3(\pmb{\theta}^{i-1}_3)$. The PS then updates $\pmb{\theta}^{i+1}$, as well as $\pmb{\theta}^{i}_1=\pmb{\theta}^{i}$ and $\pmb{\theta}^{i}_m=\pmb{\theta}^{i-1}_m$ for $m=2,3$. The next iteration $i+1$ then continues with a check of condition \eqref{con:lazy} by the PS in the same way.  

\begin{remark}\label{remark:LAG}
In the original LAG in \cite{{TGSY:18}}, instead of uploading $g_m(\pmb{\theta}^i)$, each worker $m\in\mathcal{M}_U^i$ transmits to the PS the gradient change $\fv^i_m=g_m(\pmb{\theta}^i)-g_m(\pmb{\theta}^{i-1}_m)$. The PS then estimates the full gradient $\gv(\pmb{\theta}^{i})$ by summing the vectors $\{\fv^i_m\}_{m\in\mathcal{M}_U^i}$ to the previous gradient estimate $\hat{\gv}({\pmb{\theta}^{i-1}})$ as \cite[Eq.~(6)]{TGSY:18}
\begin{align} \label{LAG:orginal}
\hat{\gv}(\pmb{\theta}^{i})&=\hat{\gv}({\pmb{\theta}^{i-1}})+\sum_{m\in\mathcal{M}_D^i}\fv_m^i =\hat{\gv}({\pmb{\theta}^{i-1}})+\sum_{m\in{\mathcal{M}^{i}_D}} \big(\gv_m(\pmb{\theta}^{i})-\gv_m(\pmb{\theta}^{i-1}_m)\big).
\end{align}
The update \eqref{update:LAG} considered here yields by direct computation the equalities 
\begin{align} \label{update:equ}
\!\!\!\!\hat{\gv}(\pmb{\theta}^{i})&=\sum_{m\in{\mathcal{M}}}\gv_m(\pmb{\theta}^{i-1}_m)+\sum_{m\in{\mathcal{M}^{i}_D}} \big(\gv_m(\pmb{\theta}^{i})-\gv_m(\pmb{\theta}^{i-1}_m)\big)\notag\\
&\stackrel{(a)}{=}\bigg(\sum_{m\in{\mathcal{M}\backslash\mathcal{M}_D^{i-1}}}\gv_m(\pmb{\theta}^{i-2}_m)+\sum_{m\in{\mathcal{M}_D^{i-1}}}\gv_m(\pmb{\theta}^{i-1})\bigg) +\sum_{m\in{\mathcal{M}^{i}_D}} \big(\gv_m(\pmb{\theta}^{i})-\gv_m(\pmb{\theta}^{i-1}_m)\big) \notag \\
&=\hat{\gv}({\pmb{\theta}^{i-1}})+\sum_{m\in{\mathcal{M}^{i}_D}} \big(\gv_m(\pmb{\theta}^{i})-\gv_m(\pmb{\theta}^{i-1}_m)\big),
\end{align}
where $(a)$ holds because we have $\pmb{\theta}^{i-2}_m=\pmb{\theta}^{i-1}_m$ for any $m\in{\mathcal{M}\backslash\mathcal{M}_D^{i-1}}$ at iteration $i-1$. From \eqref{LAG:orginal} and \eqref{update:equ}, we can see that the described LAG and the original LAG in \cite{TGSY:18} are equivalent since they use the same gradient estimate in the update \eqref{update}.
\end{remark}

%Directly from \eqref{LAG:orginal}, we have 
%\begin{align}
%\hat{\gv}(\pmb{\theta}^{i})&=\sum_{m\in{\mathcal{M}^{i}_D}}\gv_m(\pmb{\theta}^{i-1}_m)+\sum_{m\in{\mathcal{M}\backslash \mathcal{M}^{i}_D}}\gv_m(\pmb{\theta}^{i-1}_m)+\sum_{m\in{\mathcal{M}^{i}_D}} \big(\gv_m(\pmb{\theta}^{i})-\gv_m(\pmb{\theta}^{i-1}_m)\big) 
%\end{align}
%which is equal to \eqref{update:LAG}. Therefore, the described LAG here has the same gradient estimate at each iteration as the original LAG in \cite{TGSY:18}, and hence the same convergence rate. 

% Directly from \eqref{update:LAGC}, we have 
% \begin{align}
% \hat{\gv}(\pmb{\theta}^{i})&=\sum_{m\in{\mathcal{M}^{i}_D}}\gv_m(\pmb{\theta}^{i-1}_m)+\sum_{m\in{\mathcal{M}\backslash \mathcal{M}^{i}_D}}\gv_m(\pmb{\theta}^{i-1}_m)+\sum_{m\in{\mathcal{M}^{i}_D}} \big(\gv_m(\pmb{\theta}^{i})-\gv_m(\pmb{\theta}^{i-1}_m)\big) \\
% &=\bigg(\sum_{m\in{\mathcal{M}_D^{i-1}}}\gv_m(\pmb{\theta}^{i-1})+\sum_{m\in{\mathcal{M}\backslash\mathcal{M}_D^{i-1}}}\gv_m(\pmb{\theta}^{i-2}_m)\bigg)+\sum_{m\in{\mathcal{M}^{i}_D}} \big(\gv_m(\pmb{\theta}^{i})-\gv_m(\pmb{\theta}^{i-1}_m)\big),
% \end{align}

%\begin{align}
%\hat{\gv}(\pmb{\theta}^{i}) &=\sum_{m\in{\mathcal{M}}}\gv_m(\pmb{\theta}^{i-1}_m)+\sum_{m\in{\mathcal{M}^{i}_D}} \big(\gv_m(\pmb{\theta}^{i})-\gv_m(\pmb{\theta}^{i-1}_m)\big)\\
%&=\bigg(\sum_{m\in{\mathcal{M}_D^{i-1}}}\gv_m(\pmb{\theta}^{i-1})+\sum_{m\in{\mathcal{M}\backslash\mathcal{M}_D^{i-1}}}\gv_m(\pmb{\theta}^{i-2}_m)\bigg)+\bigg(\sum_{m\in{\mathcal{M}^{i}_D}} \big(\gv_m(\pmb{\theta}^{i})-\gv_m(\pmb{\theta}^{i-1}_m)\big)\bigg),
%\end{align}

\section{Lazily Aggregated Gradient Coding (LAGC)} \label{proposed}
In this section, we propose a strategy, named Lazily Aggregated Gradient Coding (LAGC), that aims at exploring the trade-off between the robustness to stragglers of GC and the computation and communication efficiency of LAG by generalizing both schemes. The idea is to cluster all the workers into groups; treat each group as a single worker in LAG; and, within each group, mitigate the effect of stragglers by applying computation redundancy as in GC. In a manner similar to LAG, the PS selects only groups of workers that have collectively large expected new contributions to the gradient. The trade-off between the robustness to stragglers of GC and the computation and communication efficiency of LAG can be controlled by selecting the size of the groups, with GC and LAG being two extreme special cases. In particular, increasing the size of the groups enhances the capability of LAGC to mitigate stragglers by utilizing storage redundancy within each group. Conversely, reducing the size of the groups gives the PS more flexibility on the selection of the subset of workers to activate at each iteration, hence potentially reducing the computation complexity.  

%As compared to GC, since less workers are active at each iteration, the communication and computation complexities can be reduced. 

% To be more precise, at each iteration $i$, the PS determines the subset of groups in a similar manner as LAG choosing the subset $\mathcal{M}_D^i$. Also similarly, for any unselected group, all the workers in the group do not need to compute. Hence, as we will see, the required communication load can be further reduced as compared to LAG. Moreover, within each group, thanks to data redundancy, the computations of $M-r+1\leq F\leq r$ workers are sufficient to recover the full gradient of assigned data for each group. It should be noted that each group can mitigate up to $r-1$ stragglers, the same as GC.

To elaborate, LAGC divides all the $M$ workers into $G=M/M_G$ groups $\mathcal{G}_1,\dots,\mathcal{G}_{G}$, each having $M_G$ workers, where design parameter $M_G$ is an integer divisor of $M$. Dataset $\mathcal{D}$ is split into $S=G$ partitions $\mathcal{D}_{1},\dots,\mathcal{D}_{G}$ of equal size, with each partition $\mathcal{D}_g$ assigned exclusively to group $\mathcal{G}_g$, for any $g\in[G]$. In each group, partition $\mathcal{D}_g$ is assigned to the $M_G$ workers in $\mathcal{G}_g$ by following GC. Accordingly, we further split $\mathcal{D}_g$ into $M_G$ equal-size batches $\mathcal{D}_{g,1}, \dots, \mathcal{D}_{g,M_G}$. Each partition is then stored at $r_G=\min\{r,M_G\}$ workers in each group. Specifically, the $m$th worker stores partition $\mathcal{D}_{g,[m+i]_{M_G}}$ for $i=0,\dots, r_G-1$. Note that, when the design parameter $M_G$ is selected as $M_G<r$, the workers' storage redundancy is underused. Furthermore, if $M_G\leq r$, all workers in a group $\mathcal{G}_g$ can fully store the partition $\mathcal{D}_g$, and hence, as in LAG and as further discussed below, no coding is needed. 
%For example, when $M_G=1$, each group consists of only one worker, which stores the entire data partition of that group without redundancy. 

%With storage capacity equal to $S_r=rS/M=r/M_G$ partitions, each worker can be assigned $r$ batches. Since we have $M_G$ batches in total, so each worker stores $b=\min\{M_G,r\}$ batches, denoted by $\mathcal{D}_{p,m(1)},\cdots,\mathcal{D}_{p,m(b)}$. This is achieved by following GC with redundancy $b$ as described in Section~\ref{sec:GC}. As a result, the effective storage redundancy in each group is given by $b$. 

%Since the storage redundancy $r$ in each group stays, the data assignment in each group can be operated by following GC. To this end, for each $p\in[P]$, we split partition $\mathcal{D}_g$ into $r$ batches $\mathcal{D}_{p,1}, \cdots, \mathcal{D}_{p,r}$. Hence there are $M$ batches in total across all the groups and each has normalized size $1/M$. With storage capacity equal to $S_r=rS/M=1$ partition of size $1/P=r/M$, each worker $m\in\mathcal{M}$ can be assigned $r$ batches, denoted by $\mathcal{D}_{p,m(1)},\cdots,\mathcal{D}_{p,m(r)}$. In each group $\mathcal{G}_{P}$, the batches are assigned in a cyclic manner, i.e., from the first worker to the last of any order, the assignment is given as $\mathcal{D}_{p,j}, \cdots, \mathcal{D}_{p,r},\cdots, \mathcal{D}_{p,j-1}$, for each $j\in[r]$. 

We denote as $\pmb{\theta}^{i}_g$ the model parameter available at all workers in group $\mathcal{G}_g$ at iteration $i$, i.e., we have $\{\pmb{\theta}^{i}_m=\pmb{\theta}^{i}_g\}_{m\in\mathcal{G}_g}$. At each iteration $i$, the PS determines the subset of groups to be activated first. To this end, the PS evaluates the condition 
\begin{align} \label{condition}
L_g^2\Big|\Big|\pmb{\theta}^{i-1}_g-\pmb{\theta}^i\Big|\Big|^2 \geq \frac{M_G^2\xi}{\alpha^2M^2D} \sum_{d=1}^{D} \Big|\Big|\pmb{\theta}^{i+1-d}-\pmb{\theta}^{i-d}\Big|\Big|^2,
\end{align}
for all groups $\{\mathcal{G}_g\}_{g=1}^{G}$, where we write $L_g$ for the smoothness constant of the local function $L_g(\pmb{\theta})=\mathop{\sum}_{\zv_n\in\mathcal{D}_g} \ell(\zv_n;\pmb{\theta})$ of each group $\mathcal{G}_g$. In a manner similar to \eqref{con:lazy}, condition \eqref{condition} is satisfied by groups that are expected to have a large new contribution to the model update \eqref{update}. This is because, the right-hand side of \eqref{condition}, by the rule \eqref{update}, represents the per-group average contribution over the most recent $D$ iterations to the approximate gradient norm squared $(1/D) \sum_{d=1}^{D}||\hat{\gv}(\pmb{\theta}^{i-d})||^2$; while, by \eqref{smooth}, the left-hand side of \eqref{condition} represents a bound on the change in the gradient squared norm expected for the local loss at group $\mathcal{G}_g$ as compared to the available gradient of the group at last iteration. 
The PS selects all the groups that satisfy condition \eqref{condition}, i.e., the subset of groups $\mathcal{I}^i=\{g\in[G]: \eqref{condition}~\text{ hold}\}$. All the workers in each group $\mathcal{G}_g$, with $g\in\mathcal{I}^i$, download the parameter $\pmb{\theta}^{i}$ from the PS, and the subset $\mathcal{M}_D^i$ of active workers is hence given as $\mathcal{M}^i_D=\bigcup_{g\in\mathcal{I}^i}\mathcal{G}_{g}$.

%\eqref{con:lazy} represents a bound on the change in the gradient squared norm expected for the local loss at worker $m$ as compared to the last available gradient from the worker. The right-hand side of \eqref{con:lazy}, by the rule \eqref{update}, represents the per-server average contribution over the most recent $D$ iterations to the approximate gradient norm squared $(1/D) \sum_{d=1}^{D}||\hat{\gv}(\pmb{\theta}^{i-d})||^2$. The PS selects the workers that satisfy condition \eqref{con:lazy}, i.e., the workers that are expected to have a sizeable difference in the local gradient, as compared to the average per-worker contribution, and hence may contribute more significantly to the model update. This yields the subset $\mathcal{M}_D^i=\{m\in\mathcal{M}: \eqref{con:lazy}~\text{holds}\}$. 

%, denoted by $\mathcal{G}_{g,f}^i$,

\begin{algorithm}[t!] 
\caption{Lazily Aggregated Gradient Coding (LAGC)}\label{alg:LAGC}
\begin{algorithmic}[1]
\State{\textbf{Input:} number of groups $G=M/M_G$, stepsize $\alpha>0$, smoothness constants $\{L_g\}_{g=1}^{G}$, number of non-straggling servers per group $F\geq M_G-r_G+1$ $\big(r_G=\min\{r,M_G\}\big)$} 
\State{\textbf{Initialize:} $\pmb{\theta}, \{\pmb{\theta}^0_g\}_{g=1}^{G}$}
\State{repeat $i=1$}
\State~~{for each group $\mathcal{G}_g$ that satisfies \eqref{condition}}
\State~~~~$\triangleright$~{all the workers in $\mathcal{G}_g$ download $\pmb{\theta}^i$ from the PS}
\State~~~~$\triangleright$~{all workers in $\mathcal{G}_{g}$ compute the gradient $\gv_g(\pmb{\theta}^{i})$}
\State~~~~$\triangleright$~{the $F$ fastest workers send GC-encoded functions $\{\fv_m^i\}$ to the PS }
\State~~{the PS recovers gradients $\{\gv_g(\pmb{\theta}^i)\}_{g\in\mathcal{I}^i}$ using GC decoding and estimates $\hat{\gv}(\pmb{\theta}^i)$ using \eqref{gra:LAGC}}
\State~~{the PS updates $\pmb{\theta}^i$ via \eqref{update} and sets $\{\pmb{\theta}^{i}_g=\pmb{\theta}^{i-1}_g\}_{g\in[G]\backslash\mathcal{I}^i}$}
\State{until convergence criterion is satisfied}
\end{algorithmic}
\end{algorithm}

\begin{algorithm}[t!] 
\caption{Grouped-Lazily Aggregated Gradient (G-LAG)}\label{alg:G-LAG}
\begin{algorithmic}[1]
\State{\textbf{Input:} number of groups $G=M/M_G$, stepsize $\alpha>0$, smoothness constants $\{L_g\}_{g=1}^{G}$, number of non-straggling servers per group $F\geq 1$ $(r_G=M_G)$} 
\State{\textbf{Initialize:} $\pmb{\theta}, \{\pmb{\theta}^0_g\}_{g=1}^{G}$}
\State{repeat $i=1$}
\State~~{for each group $\mathcal{G}_g$ that satisfies \eqref{condition}}
\State~~~~$\triangleright$~{all the workers in $\mathcal{G}_g$ download $\pmb{\theta}^i$ from the PS}
\State~~~~$\triangleright$~{all workers in $\mathcal{G}_{g}$ compute the gradient $\gv_g(\pmb{\theta}^{i})$}
\State~~~~$\triangleright$~{the fastest worker sends $\gv_g(\pmb{\theta}^{i})$ directly to the PS }
\State~~{the PS aggregates gradients $\{\gv_g(\pmb{\theta}^i)\}_{g\in\mathcal{I}^i}$ and estimates $\hat{\gv}(\pmb{\theta}^i)$ using \eqref{gra:LAGC}}
\State~~{the PS updates $\pmb{\theta}^i$ via \eqref{update} and sets $\{\pmb{\theta}^{i}_g=\pmb{\theta}^{i-1}_g\}_{g\in[G]\backslash\mathcal{I}^i}$}
\State{until convergence criterion is satisfied}
\end{algorithmic}
\end{algorithm}

For each selected group with index $g\in\mathcal{I}^i$, the PS only waits for the fastest $F$ workers to finish their computations. 
%i.e., we have $T^i_{m_{p}^i}=T^{p,i}_{r:r}$, where $T^{p,i}_{r:r}$ is the $r$th order statistic of the time variables $\{T_m^i\}_{m\in\mathcal{G}_g}$. 
As in GC, in order to guarantee the recovery of the full local gradient $\gv_g(\pmb{\theta}^{i})=\mathop{\sum}_{\zv_n\in\mathcal{D}_g} \nabla L(\zv_n;\pmb{\theta}^i)$ for each group $g$, the condition $F\geq M_G-r_G+1$ is necessary and sufficient \cite{TLDK:17}. As a result, we have the subset $\mathcal{M}^i_U=\bigcup_{g\in\mathcal{I}^i}\{m\in \mathcal{G}_g:T^i_m\leq T^{g,i}_{F:M_G}\}$ of workers uploading their gradients to the PS at iteration $i$, where $T^{g,i}_{F:M_G}$ is the $F$th order statistic of the variables $\{T_{m}^i\}_{m\in \mathcal{G}_g}$. Note, in particular, that, if $M_G= r$ yielding $r_G=M_G$, the PS can wait for the fastest worker in the group, i.e., $F=1$. In fact, each worker can compute directly the gradient $\gv_g(\pmb{\theta}^{i})$ for the partition $\mathcal{D}_g$ allocated to group $\mathcal{G}_g$.
%, which yields the subset $\mathcal{M}^i_U=\bigcup_{g\in\mathcal{I}^i}\mathcal{G}_{g,f}^i$. 
For the more general case that $M_G$ is not equal to $r$, each of the $F>1$ non-straggling workers $m$ uploads a linear combination $\fv_m^i$ of the computed partial gradients $\{\gv_{\mathcal{D}_{g,m(j)}}(\pmb{\theta}^{i})\}_{j=1}^{r_G}$ and the PS decodes the local gradient $\gv_g(\pmb{\theta}^{i})$ by following GC. 

Finally, by combining with the outdated gradients $\{\gv_g(\pmb{\theta}^{i-1}_g)\}_{g\in [G]\backslash\mathcal{I}i}$ from the inactive groups, the PS estimates the full global gradient $\gv(\pmb{\theta}^{i})$ as 
\begin{align} \label{gra:LAGC} 
\hat{\gv}(\pmb{\theta}^{i})=\sum_{g\in \mathcal{I}^i}\gv_g(\pmb{\theta}^{i})+\sum_{g\in [G]\backslash\mathcal{I}^i}\gv_g(\pmb{\theta}_g^{i-1}),
\end{align}
which is used in the update rule \eqref{update}. The PS also updates the variables $\{\pmb{\theta}^{i}_g=\pmb{\theta}^{i-1}_g\}_{g\in[G]\backslash\mathcal{I}^i}$ and $\{\pmb{\theta}^{i}_g=\pmb{\theta}^i\}_{g\in\mathcal{I}^i}$. The full algorithm is summarized in Algorithm 1. 

%This yields the subset $\mathcal{M}^i_U=\{m_{p}^i\}_{p\in\mathcal{I}^i}$. For each selected group with index $p\in\mathcal{I}^i$, the fastest worker uploads the gradient $\fv_{m_{p}^i}^i=\gv_g(\pmb{\theta}^{i})$ of the assigned data partition back to the PS. By combining with the outdated gradients $\sum_{p\in [P]\backslash\mathcal{I}i}\gv_g(\pmb{\theta}^{i-1}_g)$, the PS estimates the full global gradient $\gv(\pmb{\theta}^{i})$ as 
%\begin{align} \label{gra:LAGC}
%\hat{\gv}(\pmb{\theta}^{i}) =\sum_{p\in \mathcal{I}^i} {\fv_{m_{p}^i}^i}+\sum_{p\in [P]\backslash\mathcal{I}^i}\gv_g(\pmb{\theta}^{i-1}_g)=\sum_{p\in \mathcal{I}^i}\gv_g(\pmb{\theta}^{i})+\sum_{p\in [P]\backslash\mathcal{I}^i}\gv_g(\pmb{\theta}_g^{i-1}),
%\end{align}  
%With the estimate $\hat{\gv}(\pmb{\theta}^{i})$ in \eqref{gra:LAGC}, $\pmb{\theta}^{i}$ is updated with the rule in \eqref{update}. The PS also updates the stored information $\{\pmb{\theta}^{i}_g=\pmb{\theta}^{i-1}_g\}_{p\in[P]\backslash\mathcal{I}^i}$ and $\{\pmb{\theta}^{i}_g=\pmb{\theta}^i\}_{p\in\mathcal{I}^i}$.

\begin{remark}
%We can observe that, for LAGC, the choice of $M_G$ decides the communication and computation complexities. By means of grouping, each group is able to tolerant up to $b-1$ stragglers thanks to the storage redundancy $b$. It should be noted that, when we have $M_G\geq r$, i.e., $b=r$, the stragglers tolerable in each group can be as many as in GC, given as $r-1$. As discussed, LAGC generalizes both LAG and GC by choosing the corresponding $M_G$. In particular, when we choose $M_G=M$, and hence we have $b=r$, LAGC is reduced to GC where there is only one group consisting of all the workers. While when we choose $M_G=1$, and hence we have $b=1$, LAGC becomes LAG where there are $M$ groups with one worker in each group. Hence, as will be shown later, LAGC consumes less time each iteration and converges much faster in terms of wall clock time, as compared to LAG and GC both. 
As discussed, LAGC generalizes both LAG and GC: when choosing $M_G=M$ and $\xi=0$, LAGC reduces to GC, while setting $M_G=1$ recovers LAG. Intermediate values of $M_G$ yield novel schemes. 
\end{remark}

\begin{remark} \label{GLAG}
Setting the number of groups to be smaller or equal to the storage redundancy, i.e., $M_G\leq r$, yields a novel scheme that does not require coding within each group, while still benefiting from both robustness to stragglers and reduced computation complexity. With this choice, each worker $m\in\mathcal{G}_g$ stores the entire data partition $\mathcal{D}_g$ for group $g\in[G]$. Hence, for each selected group $\mathcal{G}_g$ with $g\in\mathcal{I}^i$, the PS only needs to wait for the fastest worker (i.e., $F=1$), since the latter can send the desired gradient $\fv_m^i=\gv_g(\pmb{\theta}^{i})$ directly to the PS. To highlight the fact that no coding is involved, we refer to this set of schemes as Grouped-LAG (G-LAG). We note that setting $\xi=0$, and hence selecting all groups at all times, G-LAG reduces to Grouped-GD (G-GD) \cite{OFU:19}.  
%As compared to the rest cases, grouped LAG is able to tolerant more stragglers, given as $M(r-1)/r$, since there are $M/r$ groups and each group can tolerant $b-1=r-1$ stragglers. For the cases $M_G>r$, the maximum number of stragglers is $M(r-1)/M_G$, while for $M_G<r$ whereby the storage of the workers is underused, it is $M(M_G-1)/M_G$. As an example, from Fig.~\ref{fig:group}(a) and (b), which corresponds to $M_G>r=2$ and $M_G=r$, we can observe that the former can have two stragglers at most, with one from each group, while the latter can have three at most. 
\end{remark}

%\begin{remark}
%As described for GC, coding is able to reduce the communication complexity by combining the computed partial gradients at each active worker in each iteration. By using the same idea for each group, we can see that LAGC with the options $M_G>r$ also has the communication-efficient advantage as compared to G-LAG which does not requiring coding.
%\end{remark}

%We use $\pmb{\theta}^{t}_{m}$ to denote the version of $\pmb{\theta}$ at iteration $t$ stored by worker $m$. For each worker $m\in\mathcal{G}_i$, with any $i\in\mathcal{M}_{g}^t$, we have $\pmb{\theta}^{t}_{m}=\pmb{\theta}^{t}$ from the PS, otherwise, we have $\pmb{\theta}^{t}_{m}=\pmb{\theta}^{t-1}_m$.

%Since all workers perform the computations with global gradient $\gv(\pmb{\theta}^{t})$, we use $\pmb{\theta}^{t}_{m}$ to denote the stored old copy. 

 %\begin{figure}[t!] 
  %\centering
%\includegraphics[width=1\columnwidth]{groupa.pdf}
 %\caption{LAGC with two different grouping strategies: (a) $M_G=2$ and (b) $M_G=3$.}
%\label{fig:group}
%\end{figure}

\begin{figure}
\centering
\subfigure{%
\includegraphics[width=0.35\textwidth]{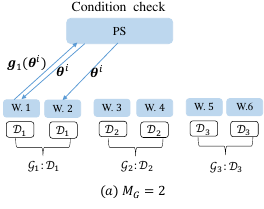}}%
\qquad
\subfigure{%
\includegraphics[width=0.35\textwidth]{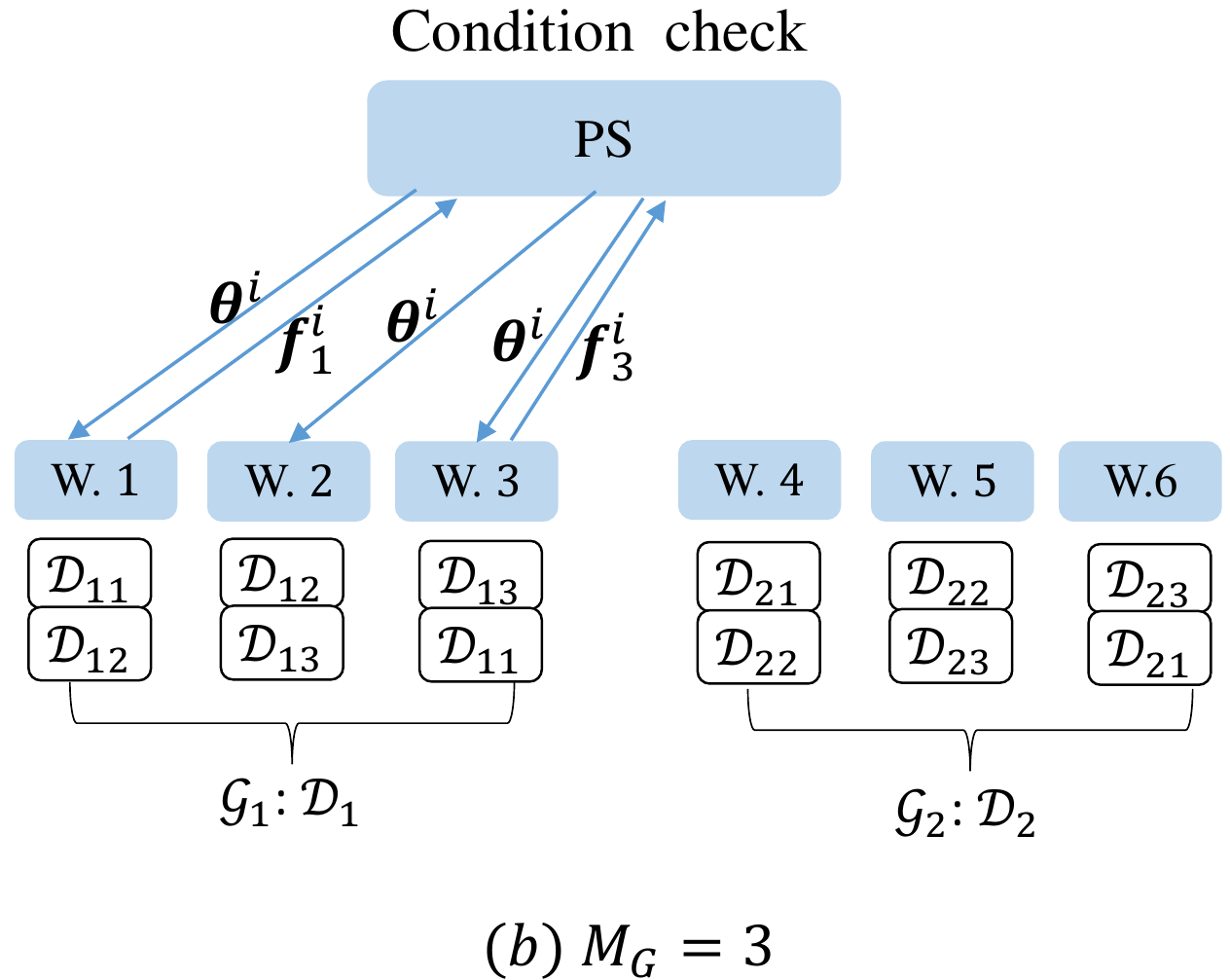}}%
\caption{LAGC with two different grouping strategies: (a) $M_G=2$ and (b) $M_G=3$.}
\label{fig:group}
\end{figure}

\emph{Example:} Consider $M=6$ workers and storage redundancy $r=2$. In this case, $M_G$ can take values $M_G=2$ or 3, apart from the cases $M_G=1$ and $6$, corresponding to LAG and GC, respectively. For $M_G=2$, we have the G-LAG scheme described in Remark \ref{GLAG}, whereby each worker in each group can store the entire data partition of that group and no coding is necessary, as shown in Fig.~\ref{fig:group}(a). In the example of Fig.~\ref{fig:group}(a), the PS selects only group $\mathcal{G}_1$ using condition \eqref{condition}, and the PS obtains $\gv_1(\pmb{\theta}^i)=\fv_1^i$ directly from the fastest worker in $\mathcal{G}_1$. For $M_G=3$, the workers are clustered into two groups and the partition for each group is divided into two parts, each stored at two workers with redundancy $r_G=2$, as shown in Fig.~\ref{fig:group}(b). In the illustration of Fig.~\ref{fig:group}(b), only group 1 satisfies condition \eqref{condition}. Hence, worker $1,2$ and 3 in group $\mathcal{G}_1$ download the model $\pmb{\theta}^i$ from the PS, and the PS waits for the fastest $F=2\geq M_G-r_G+1$ workers to finish their computations. GC is used to recover the gradient $\gv_1(\pmb{\theta}^i)$ from group $\mathcal{G}_1$. By summing up the outdated partial $\gv_2(\pmb{\theta}^{i-1}_2)$, the PS estimates the global gradient as $\hat{\gv}(\pmb{\theta}^i)=\gv_1(\pmb{\theta}^i)+\gv_2(\pmb{\theta}^{i-1}_2)$.

\section{Analysis} \label{analysis}
In this section, we analyze the \emph{time complexity} \eqref{def:time}, \emph{communication complexity} \eqref{def:commu}, and \emph{computation complexity} \eqref{def:compu} of all the schemes considered above, which are summarized in Table \ref{tech}. Note that, as a reference, we also study standard Gradient Descent (GD), which is a special case of GC without data redundancy, i.e., with $r=1$; as well as G-GD, which, as seen, is an extreme case of G-LAG where all the groups are active at each iteration. The results of the analysis in this section are summarized in Section~\ref{illustra} via numerical illustrations.

In order to derive the mentioned metrics, we first discuss the iteration complexity $I_{\epsilon}$ in \eqref{comple:iter}. A standard result is that GD has iteration complexity 
\begin{align}
I_{\epsilon}^{GD}=\bar{I}_{\epsilon}=\kappa\log\Big(\frac{\Delta L^0}{\epsilon}\Big),
\end{align} 
with $\kappa=L/\mu$ being the condition number for the training empirical loss \eqref{loss} and $\Delta L^0=L(\pmb{\theta}^{0})-L(\pmb{\theta}^*)$ being the difference between the loss $L(\pmb{\theta}^{0})$ at the initial iteration $\pmb{\theta}^{0}$ and the loss at the optimal point $\pmb{\theta}^*$ \cite{MS}. We note that this result is obtained by choosing a stepsize $\alpha<1/L$. By constructions, GC and G-GD have the same iteration complexity $I_{\epsilon}^{GC}=I_{\epsilon}^{G-GD}=\bar{I}_{\epsilon}$. For LAG, as shown in \cite{TGSY:18}, by choosing the stepsize as $\alpha<1/L$, we have the iteration complexity 
\begin{align}
I_{\epsilon}^{LAG}=\frac{\bar{I}_{\epsilon}}{\alpha L},
\end{align} 
which shows the same scaling with $\epsilon$ and $\kappa$ as GD. Finally, for LAGC, the iteration complexity is by construction equivalent to that of LAG with $G$ workers. In fact, since LAGC treats each group of workers as a single worker, its operation
is equivalent to a system with a smaller number of workers. Given that the iteration complexity of LAG does not depend on the number of workers, we have $I_{\epsilon}^{LAGC}=I_{\epsilon}^{LAG}=\bar{I}_{\epsilon}/(\alpha L)$.

%For LAGC, under Assumption 1 and 2, to achieve $\epsilon$-optimality gap with the chosen stepsize $\alpha=(1-\sqrt{\xi})/L$, the required number of iterations is given as 
%\begin{align}
%I_{LAGC}=\frac{\kappa}{1-\sqrt{\xi}}\log(\epsilon^{-1}).
%\end{align}  

%\begin{proof}
%At each iteration, LAGC selects a subset of $G$ groups and each selected group computes the partial gradient of the assigned data partition. By taking each group as a single worker in LAG, LAGC is equivalent to LAG with $G$ workers in total with respect to the gradient estimate, i.e., the iteration complexity. Hence, by following the same analysis for the iteration complexity LAG \cite{TGSY:18}, we can obtain the same iteration complexity for LAGC. Hence, similar to LAG, the iteration complexity of LAGC is also on the same order of GC. 
%\end{proof}

\begin{table} 
\caption{Summary of techniques considered in this worker}
\begin{center}
  \begin{tabular}{ | c || c | c| c| } 
    \hline
                 & \text{Coding} & \text{Adaptive selection} & \text{Grouping} \\ \hline\hline
    \text{GD}    & $\times$      & $\times$       & $\times$ \\ \hline
    \text{GC \cite{TLDK:17}}    & $\checkmark$  & $\times$       & $\times$  \\ \hline
	  \text{LAG \cite{TGSY:18}}   & $\times$      & $\checkmark$   & $\times$  \\ \hline
	  \text{G-GD \cite{OFU:19}}  & $\times$      & $\times$       & $\checkmark$  \\ \hline
	  \text{G-LAG [this paper]} & $\times$      & $\checkmark$   & $\checkmark$  \\ \hline
	   \text{LAGC [this paper]} & $\checkmark$  & $\checkmark$   & $\checkmark$  \\ \hline
  \end{tabular}
	\label{tech}
\end{center}
\end{table}

\subsection{Wall-clock Time Complexity}

We proceed to analyze the average wall-clock time complexity \eqref{def:time}. To elaborate, we define as $T_{a:b}$ the $a$th order statistics of i.i.d. variables $\{T_i\}_{i=1}^b$, that is, the $a$th smallest value in the set $\{T_i\}_{i=1}^b$, where $a$ and $b$ are two integers satisfying $a\leq b$. The average $\bar{T}_{a:b}=\mathrm{E}[T_{a:b}]$ can be written in closed form for the two representative distributions considered here. In particular, we have 
\begin{align}\label{av:expo}
\bar{T}_{a:b}(r):=\mathrm{E}[T_{a:b}]=\eta r(H_b-H_{b-a})
\end{align}
for the case of exponential distribution with mean $\eta r$ \cite{MCJ:18}, where $H_a=\sum_{k=1}^{a}1/k$ is the $a$th harmonic number; and 
\begin{align}\label{av:pareto}
\bar{T}_{a:b}(r)&:=\mathrm{E}[T_{a:b}] =\frac{\eta r (1-\beta)}{\beta} \frac{\Gamma (b-a+1-1/\beta)\Gamma (n+1)}{\Gamma (b-a+1)\Gamma (b+1-1/\beta)}
\end{align}
for the case of Pareto distribution with scale-shape pair $(\eta r(\beta-1)/\beta,\beta)$ with $\beta>1$, where $\Gamma(x)$ is Gamma function given by $\Gamma(x)=\int_{0}^{\infty}t^{x-1}e^{-t}dt$ \cite{V:76}. Note that both averages \eqref{av:expo} and \eqref{av:pareto} increase with the mean $\eta r$ of each variable $T_i$ and with $b$. To ease the notation, we also write $\bar{T}_a(r)=\bar{T}_{a:a}(r)$ in the following.

\emph{GD:} The average run-time of each iteration for GD is given as $\bar{T}_M(1)$, since the PS waits for all $M$ servers to complete their computations at each iteration and no computational redundancy is leveraged, i.e., $r=1$. This yields the overall run-time
\begin{align}  \label{time:GD}
\bar{T}_{\epsilon}^{GD}=\bar{I}_{\epsilon}\bar{T}_M(1).
\end{align} 
 
\emph{GC:} With GC, at each iteration $i\in I_{\epsilon}^{GC}$, the PS waits only for the set $\mathcal{M}_U^i$ of the fastest $F\geq M-r+1$ workers to finish their computations, yielding the average run-time $\bar{T}_{F:M}$. As a result, the overall run-time is given as  
%\begin{align}  \label{time:GC}
%\bar{T}_{\epsilon}^{GC}=\mathrm{E}\bigg[\sum_{i=1}^{I_{GC}}\max_{m\in\mathcal{M}_U^i}\{T_m^i\}\bigg]=\sum_{i=1}^{I_{GC}}\mathrm{E}\Big[\max_{m\in\mathcal{M}_U^i}\{T_m^i\}\Big]=\sum_{i=1}^{I_{GC}}\mathrm{E}[T_{F:M}]=I_{GC}\frac{H_M-H_{M-F}}{\eta S_r}.
%\end{align}
\begin{align}  \label{time:GC}
\bar{T}_{\epsilon}^{GC}=\bar{I}_{\epsilon}\bar{T}_{F:M}(r).
\end{align} 
Comparing with \eqref{time:GD}, GC can reduce the time complexity if $\bar{T}_{F:M}(r)<\bar{T}_{M}(1)$.

\emph{LAG:} At each iteration $i\in I_{\epsilon}^{LAG}$ of the LAG scheme which assumes $r=1$, all the selected workers in subset $\mathcal{M}_U^i$ have to complete their computations, yielding the average run-time $\bar{T}_{|\mathcal{M}_U^i|}(1)$. From \cite[Lemma 4]{TGSY:18}, given an integer $d\in\{0,1,\dots,D\}$, a worker with smoothness parameter $L_m$ satisfying the inequalities
\begin{align} \label{num:sat}
\bar{L}_{d+1}^2<L_m^2<\bar{L}^2_{d}, ~~\text{with}~\bar{L}_{d}^2=\frac{\xi}{Dd\alpha^2M^2},
\end{align}
is selected in at most $I_{\epsilon}^{LAG}/(d+1)$ iterations. A larger smoothness constant, and hence a less sensitive gradient, cause a worker to be selected less frequently. Using this result, we can bound the average number $(1/I_{\epsilon}^{LAG})\sum_{i=1}^{I_{\epsilon}^{LAG}}|\mathcal{M}_U^i|$ of selected servers per iteration as
\begin{align} \label{eq:boundLAG}
\frac{1}{I_{\epsilon}^{LAG}}\sum_{i=1}^{I_{\epsilon}^{LAG}}|\mathcal{M}_U^i|\leq M\sum_{d=0}^{D}\frac{h(d)}{d+1}:=\bar{M}\leq M,
\end{align}
where we have defined the function $h(d)=(1/M)\sum_{m\in\mathcal{M}}\mathds{1}(\bar{L}_{d+1}^2<L_m^2<\bar{L}^2_{d})$ with $\bar{L}_{0}=\bar{L}_{D+1}=0$. Note that function $h(d)$ indicates the fraction of the workers satisfying condition \eqref{num:sat}. Therefore, parameter $\bar{M}$ increases when all smoothness constant $\{L_m\}$ decrease. From \cite{TGSY:18}, we have that $\sum_{d=0}^{D}h(d)/(d+1)\leq 1$, i.e., $\bar{M}\leq M$. As proved in Appendix~\ref{sec:proof}, the bound \eqref{eq:boundLAG} can be used in turn to bound the overall time as
%for $d=a2,\cdots,D$, $h(0)=1$ and $h(D+1)=0$ \cite{TGSY:18}. That is, through $I_{\epsilon}^{LAG}$ iterations, we have $M(h(d)-h(d+1))$ workers out of $M$ workers that communicate to the PS $I_{\epsilon}^{LAG}/(d+1)$ iterations, for any $d=0,\cdots, D$. 
%$\mathrm{E}[\max_{m\in\mathcal{M}_U^i}\{T_m^i\}]=H_{|\mathcal{M}_U^i|}/(\eta S_r)$ by using the $|\mathcal{M}_U^i|$th order statistic of variables $\{T_m^i\}_{m\in\mathcal{M}_U^i}$. As a result, the overall run-time is given as 
\begin{align}  \label{time:LAG}
\bar{T}_{\epsilon}^{LAG}=\sum_{i=1}^{I_{\epsilon}^{LAG}}\bar{T}_{|\mathcal{M}_U^i|}\leq \frac{\bar{I}_{\epsilon}}{\alpha L} \bar{T}_{\bar{M}}(1).
%I_{\epsilon}^{LAG} \bar{T}_{\frac{\bar{M}_{U}^{LAG}}{I_{\epsilon}^{LAG}}}.
\end{align}
Comparing with \eqref{time:GD}, we see that LAG can only decrease the time complexity as compared to GD if $\alpha\approx 1/L$ and the smoothness constant $\{L_m\}$ are large. 
%where we have defined $M_{LAG}=(\sum_{i=1}^{I_{\epsilon}^{LAG}}|\mathcal{M}_U^i|)/I_{\epsilon}^{LAG}\leq\bar{C}_{D,max}^{LAG}/I_{\epsilon}^{LAG}$. To be more precise, when the sum $\sum_{i=1}^{I_{\epsilon}^{LAG}}|\mathcal{M}_U^i|$ is uniformly distributed over all iterations, the time complexity takes the maximum. The detailed proof is presented in Appendix~\ref{sec:proof}. 
%Since the run-timeis equal to the maximum of the selected $|\mathcal{M}_U^i|$ workers, which is given as $\mathrm{E}\big[\max_{m\in\mathcal{M}_U^i}\{T_m^i\}\big]=H_{|\mathcal{M}_U^i|}/(\eta S_r)$.

\emph{G-GD:} Based on grouping, at each iteration $i$, the PS waits for the fastest $F\geq M_G-r_G+1$ workers in each group to finish their computations. Hence, the run-time $T_g^G$ for each group is the $F$th order statistic of the random times $\{T_i\}_{i\in\mathcal{G}_g}$ of the workers in group $\mathcal{G}_g$. In a manner similar to \eqref{av:expo}-\eqref{av:pareto}, we define $\bar{T}_{a:M_G}^G(r)$ to be the expectation of the $a$th order statistics of the random variables $\{T_g^G\}_{g=1}^{M_G}$, and we denote $\bar{T}_{M_G:M_G}^G(r)=\bar{T}_{M_G}^G(r)$. Unlike \eqref{av:expo}-\eqref{av:pareto}, this expectation does not generally have a closed form, but it can be computed numerically as
\begin{align}
\bar{T}^G_{a:M_G}(r)=\int_{0}^{+\infty} \Big(1-\big(F^G(x)\big)^{a}\Big)dx,
\end{align}
where we have defined the Cumulative Distribution Function (CDF) $F^G(x)=\sum_{j=F}^{M_G}\binom{M_G}{j}(F(x))^j(1-F(x))^{M_G-j}$ of each variable $T_g^G$ \cite{CBN:08}. With these definitions, the overall run-time of G-GD is given as  
\begin{align}  \label{time:G-GD}
\bar{T}_{\epsilon}^{G-GD}=I_{\epsilon}^{G-GD}\bar{T}_{M_G}^G=\bar{I}_{\epsilon}\bar{T}_{M_G}^G.
\end{align} 
Based on \eqref{time:G-GD}, G-GD can reduce the time complexity if $\bar{T}_{M_G}^G(r)<\bar{T}_M(r)$.

\emph{LAGC:} At each iteration $i\in I_{\epsilon}^{LAGC}$, LAGC selects a subset $\mathcal{I}^i$ of groups $\mathcal{G}_g$ with $g\in\mathcal{I}^i$. Using the same approach discussed above from \cite{TGSY:18}, the average group size $(1/I_{\epsilon}^{LAGC})\sum_{i=1}^{I_{\epsilon}^{LAGC}}|\mathcal{I}^i|$ can be upper bounded as
\begin{align} \label{eq:boundLAGC}
\frac{1}{I_{\epsilon}^{LAGC}}\sum_{i=1}^{I_{\epsilon}^{LAGC}}|\mathcal{I}^i|\leq G\sum_{d=0}^{D}\frac{h_G(d)}{d+1}:=\bar{G}\leq G,
\end{align}
where we have defined the function $h_G(d)=(1/G)\sum_{g\in[G]}\mathds{1}(\bar{L}_{G,d+1}^2<L_g^2<\bar{L}^2_{G,d})$, with $\bar{L}_{G,d}^2=\xi/(Dd\alpha^2G^2)$ and $\bar{L}_{G,0}=\bar{L}_{G, D+1}=0$. This distribution has the same interpretation given above for function $h(d)$, with groups replacing workers. 
Similarly, we have that $\sum_{d=0}^{D}h_G(d)/(d+1)$, i.e., $\bar{G}\leq G$.
To allow all the selected groups in subset $\mathcal{I}^i$ to complete the computations, the average run-time of each iteration is given as $\bar{T}^G_{|\mathcal{I}^{i}|:M_G}(r)$. As a result, the overall average run-time can be upper bounded as 
\begin{align}  \label{time:LAGC}
\bar{T}_{\epsilon}^{LAGC}=\sum_{i=1}^{I_{\epsilon}^{LAGC}}\bar{T}^G_{|\mathcal{I}^{i}|}\leq \frac{\bar{I}_{\epsilon}}{\alpha L}\bar{T}^G_{\bar{G}}(r).
%\bar{T}_{\epsilon}^{LAGC}=\sum_{i=1}^{I_{\epsilon}^{LAGC}}\bar{T}^G_{|\mathcal{I}^{i}|}\leq I_{\epsilon}^{LAGC}\mathrm{E}\bigg[T^G_{\frac{\bar{G}^{LAGC}}{I_{\epsilon}^{LAGC}}:{\frac{\bar{G}^{LAGC}}{I_{\epsilon}^{LAGC}}}}\bigg],
\end{align}
Comparing with \eqref{time:GD}, LAGC can hence outperform GD in terms of time complexity if $\bar{T}^G_{\bar{G}}(r)/(\alpha L)\leq \bar{T}_M(1)$.

%To prove \eqref{time:LAGC}, we consider $M$ random variables $\{T_i\}_{i=1}^{M}$. We then split them into $G$ groups $\mathcal{G}_1,\cdots,\mathcal{G}_G$, with each having $M_G=M/G$ variables, denoted by $\{T_i^g\}_{i\in[M_G]}$. Furthermore, we define as $T_{p:M_G}^g$ being the $p$th order statistic of each group, where integer $p$ satisfying $p\leq M_G$. The CDF of each variable $T_{p:M_G}^g$ is hence given as $F_G(x)=\sum_{j=p}^{M_G}\binom{M_G}{j}(F(x))^j(1-F(x))^{M_G-j}$ \cite{CBN:08}.
%It is easy to see that the variables $\{T_{p:M_G}^g\}_{g\in[G]}$ are i.i.d. due to the independence of the variables $\{T_i\}_{i=1}^{M}$. Thus, by following \cite[Lemma 2]{HAD:97}, we again have the equality 
%\begin{align} \label{appen:lagc1}
%\mathrm{E}[T^G_{a:a}]-\mathrm{E}[T^G_{a-1:a-1}]=\int^{+\infty}_{-\infty} F_G^{a-1}(x)[1-F_G(x)]dx.
%\end{align}
%where $T^G_{a:a}$ is the largest order statistic of any $a$ variables from $\{T_{p:M_G}^g\}_{g\in[G]}$ with $a\leq G$. Hence, $f^G(a)=E[T^G_{a:a}]$ is discrete concave, which satisfies $f^G(S_a/K)\geq 1/K\sum_{a=1}^{K}f^G(a)$. Similar to LAG, by setting $a_i=|\mathcal{I}^i|$ and $K=|\mathcal{I}_{\epsilon}^{LAGC}|$, we can have the desired inequality in \eqref{time:LAGC}. 

\subsection{Communication Complexity}
We proceed to investigate the communication complexity.

\emph{GD:} With the conventional GD, the overall communication load is given as
\begin{align}  \label{commu:GD}
\bar{C}_{\epsilon}^{GD}=\mathrm{E}\bigg[\sum_{i=1}^{I_{\epsilon}^{GD}}|\mathcal{M}_D^i|+|\mathcal{M}_U^i|\bigg]=2M\bar{I}_{\epsilon},
\end{align}
since all $M$ workers download and upload model parameters from the PS.
%, and each worker uploads the local gradient of the same size as the global gradient $\gv(\pmb{\theta}^{i})$, i.e., with $\lambda=1$. 

\emph{GC:} At each iteration $i\in I_{\epsilon}^{GC}$ of GC, there are $|\mathcal{M}^i_D|=M$ workers downloading the model parameter and $|\mathcal{M}^i_U|=F$ workers uploading the computed results.
%, which has the same size of the global gradient $\gv(\pmb{\theta}^{i})$ by using coding, i.e., $\lambda=1$. 
Hence, the overall communication load is given as
\begin{align} \label{commu:GC}
\bar{C}_{\epsilon}^{GC}=\mathrm{E}\bigg[\sum_{i=1}^{I_{\epsilon}^{GC}}|\mathcal{M}_D^i|+|\mathcal{M}_U^i|\bigg]=\bar{I}_{\epsilon}(M+F).
\end{align}
Comparing \eqref{commu:GC} with \eqref{commu:GD}, we observe that GC reduces the communication complexity by a factor $2/(1+F/M)$, which can be as large as 2 if $F/M$ is small enough. 

\emph{LAG:} At each iteration $i\in I_{\epsilon}^{LAG}$ of LAG, all the workers in subset $\mathcal{M}^i_D$ download from the PS and also upload model parameters to the PS. Hence, the overall communication load is given as
\begin{align}  \label{commu:LAG}
\bar{C}_{\epsilon}^{LAG}&=\mathrm{E}\bigg[\sum_{i=1}^{I_{\epsilon}^{LAG}}|\mathcal{M}_D^i|+|\mathcal{M}_U^i|\bigg] =2\mathrm{E}\bigg[\sum_{i=1}^{I_{\epsilon}^{LAG}}|\mathcal{M}_U^i| \bigg]  \leq 2\bar{M}\frac{\bar{I}_{\epsilon}}{\alpha L},
%&\leq 2MI_{\epsilon}^{LAG}\sum_{d=0}^{D}\frac{1}{d+1}\big(h(d)-h(d+1)\big)= 2MI_{\epsilon}^{LAG}\bigg(1-\sum_{d=1}^{D}\Big(\frac{1}{d}-\frac{1}{d+1}\Big)h(d)\bigg):=2\bar{C}_{D,max}^{LAG},
\end{align}
with $\bar{M}$ defined in \eqref{eq:boundLAG}. Since we have $\bar{M}\leq M$, LAG can reduce the communication complexity as long as $\alpha L$ is not too small. 
%the smoothness constants $\{L_m\}$ are large so that $\bar{M}$ is smaller than $M$. 
%where we have the indicator function $h(d)=(1/M)\sum_{m\in\mathcal{M}}\mathds{1}\big(L_m^2\leq \xi/(Dd\alpha^2M^2)\big)$ for $d=2,\cdots,D$, $h(0)=1$ and $h(D+1)=0$ \cite{TGSY:18}. Through $I_{\epsilon}^{LAG}$ iterations, we have $M(h(d)-h(d+1))$ workers out of $M$ workers that communicate to the PS $I_{\epsilon}^{LAG}/(d+1)$ iterations, for any $d=0,\cdots, D$. As compared to the computation complexity in \eqref{commu:GD}, $1-\sum_{d=1}^{D}\big(1/d-1/(d+1)\big)h(d)$ from \eqref{commu:LAG} represents the fraction of the reduced communication per iteration. 

\emph{G-GD:} With G-GD, the overall communication load within $I_{\epsilon}^{G-GD}$ iterations is given as
\begin{align}  \label{commu:GD}
\bar{C}_{\epsilon}^{G-GD}&=\mathrm{E}\bigg[\sum_{i=1}^{I_{\epsilon}^{G-GD}}|\mathcal{M}_D^i|+|\mathcal{M}_U^i|\bigg] =(M+G(M_G-r_G+1))\bar{I}_{\epsilon},
\end{align}
since all $M$ workers download the model parameters but only the fastest $M_G-r_G+1$ workers in each group upload the computed results. Hence, G-GD reduces the communication complexity as compared to GD by a factor that increases with the computation redundancy factor $r_G$.

%\begin{figure}
%\centering
%\subfigure[Exponential distribution]{%
%\label{fig:firscommu}%
%\includegraphics[width=0.44\textwidth]{commu_complexity_ex.pdf}}%
%\qquad
%\subfigure[Pareto distribution]{%
%\label{fig:secondcommu}%
%\includegraphics[width=0.46\textwidth]{commu_complexity_pa.pdf}}%
%\caption{Communication complexity $\bar{C}_\epsilon$ of GC, LAG and LAGC versus wall-clock time complexity for $M=16$, $r=4$, $\eta=0.05$, $\alpha=1.1$, $\xi=1$, and $L_m=(1.3^{m-1}+1)^2$, for any $m\in[M]$.}
%\label{fig:commu_complexity}
%\end{figure}

\emph{LAGC:} At each iteration $i\in I_{LAGC}$ of LAGC, $|\mathcal{I}^i|$ groups of workers are chosen to download the model parameter, which amount to $|\mathcal{M}_D^i|=|\mathcal{I}^i|M_G$. In each group, $F\geq M_G-r_G+1$ workers upload their computations, yielding $|\mathcal{M}_U^i|=|\mathcal{I}^i|F$. As a result, the overall communication complexity within $I_{LAG}$ iterations is given as  
\begin{align} \label{comm:LAGC}
\bar{C}_{\epsilon}^{LAGC}&=\mathrm{E}\bigg[\sum_{i=1}^{I_{\epsilon}^{LAGC}}|\mathcal{M}_D^i|+|\mathcal{M}_U^i|\bigg] =\mathrm{E}\bigg[\sum_{i=1}^{I_{\epsilon}^{LAGC}}|\mathcal{I}^i|(M_G+F)\bigg] \leq (M_G+F)\bar{G}\frac{\bar{I}_{\epsilon}}{\alpha L},
\end{align}
with $\bar{G}$ defined in \eqref{eq:boundLAGC}. Since the average number $M_G\bar{G}\leq M_G G=M$ of selected workers for download and the number $F\bar{G}\leq M$ of selected workers for upload are both no larger than the total number $M$ of workers, LAGC can outperform GD in terms of communication complexity when $\alpha L$ is close to one.

\subsection{Computation Complexity}
Finally, we evaluate the computation complexity \eqref{def:compu}.

\emph{GD:} The overall computation complexity over $I_{\epsilon}^{GD}$ iterations for GD is given as
\begin{align} 
\bar{P}_{\epsilon}^{GD}=\mathrm{E}\bigg[\sum_{i=1}^{I_{\epsilon}^{GD}}\frac{1}{M}|\mathcal{M}_D^i|\bigg]=\bar{I}_{\epsilon},
\end{align}
since the gradient for each data point is computed as each iteration. 

\emph{GC:} With a data redundancy $r\leq 1$ at the workers, the overall computation complexity of GC is given as
\begin{align} 
\bar{P}_{\epsilon}^{GC}=\mathrm{E}\bigg[\sum_{i=1}^{I_{\epsilon}^{GC}} \frac{r}{M}|\mathcal{M}_D^i|\bigg]=r\bar{I}_{\epsilon}.
\end{align}
The computation load of GC is hence $r$ times that of GD. 

\emph{LAG:} The overall computation complexity of LAG is similarly bounded as
\begin{align}  \label{compu:LAG}
\bar{P}_{\epsilon}^{LAG}=\mathrm{E}\bigg[\sum_{i=1}^{I_{\epsilon}^{LAG}}\frac{1}{M}|\mathcal{M}_D^i|\bigg]\leq \frac{\bar{M}}{M}\frac{\bar{I}_{\epsilon}}{\alpha L}.
\end{align}
Therefore, LAG can reduce the computation complexity if $\alpha L$ is close to one. 
%the average number $\bar{M}$ of selected workers is sufficiently smaller than $M$.

\emph{G-GD:} The overall computation complexity of G-GD is given as
\begin{align}  \label{compu:G-GD}
\bar{P}_{\epsilon}^{G-GD}=\mathrm{E}\bigg[\sum_{i=1}^{I_{\epsilon}^{G-GD}}\frac{r_G}{M}|\mathcal{M}_D^i|\bigg]=r_G\bar{I}_{\epsilon},
\end{align}
yielding an increase equal to $r_G$ in computation complexity.

\emph{LAGC:} Finally, the overall computation complexity of LAGC is bounded as  
\begin{align} \label{compu:LAGC}
\bar{P}_{\epsilon}^{LAGC}=\mathrm{E}\bigg[\sum_{i=1}^{I_{\epsilon}^{LAGC}}\frac{r_G}{M}|\mathcal{M}_D^i|\bigg]\leq\frac{r_G M_G \bar{G}}{M} \frac{\bar{I}_{\epsilon}}{\alpha L}.
\end{align}
since we have the inequality $M_G \bar{G}/M\leq 1$ due to adaptive selection, LAGC can reduce the computation complexity of GC when the average number $\bar{G}$ of groups satisfies the inequality $\bar{G}\leq M^2/( M_G r_G)$ as long as $\alpha L \approx 1$.
%
 %$r_G M_G \bar{G}/M\leq M$, i.e., the saving from selection is larger than the cost from the redundancy.  

%\begin{figure}[t!] 
  %\centering
%\includegraphics[width=0.45\columnwidth]{compu_complexity.pdf}
%\caption{Computation complexity $\bar{P}_\epsilon$ of GC, LAG and LAGC that guarantees the $\epsilon$-optimality gap with $\epsilon=10^{(-8)}$, where LAGC has five different grouping strategies depending on the value of $M_G$ under the consideration of $M=16$ and $r=4$.}
%\label{fig:compu_complexity}
%\end{figure}

%\begin{figure}%
%\centering
%\subfigure[Exponential distribution]{%
%\label{fig:firstcompu}%
%\includegraphics[width=0.43\textwidth]{compu_complexity_ex.pdf}}%
%\qquad
%\subfigure[Pareto distribution]{%
%\label{fig:secondcompu}%
%\includegraphics[width=0.46\textwidth]{compu_complexity_pa.pdf}}%
%\caption{Computation complexity $\bar{P}_\epsilon$ of GC, LAG and LAGC versus wall-clock time complexity for $M=16$, $r=4$, $\eta=0.05$, $\alpha=1.1$, $\xi=1$, and $L_m=(1.3^{m-1}+1)^2$, for any $m\in[M]$.}
%\label{fig:commu_complexity}
%\label{fig:compu_complexity}
%\end{figure}

\subsection{Numerical Illustration} \label{illustra}

\begin{figure}[t!] 
  \centering
\includegraphics[width=0.6\columnwidth]{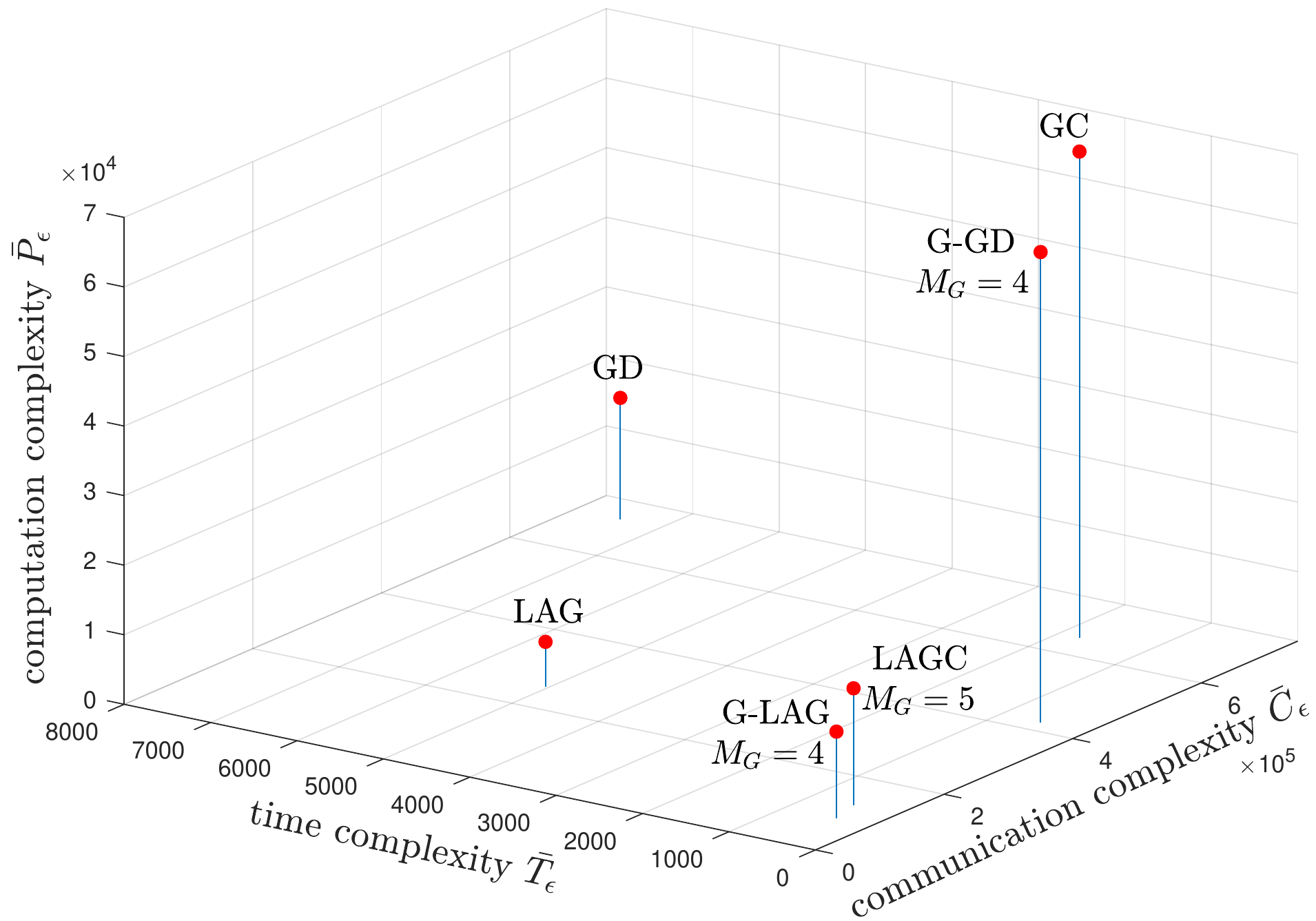}
\caption{Time, communication, and computation complexity measures for GD, GC, LAG and LAGC that guarantees the $\epsilon$-optimality gap with $\epsilon=10^{-8}$ under the Pareto distribution for the workers' computing times. }
\label{fig:complexity_pa_3D}
\end{figure}

\begin{figure}[t!] 
  \centering
\includegraphics[width=0.6\columnwidth]{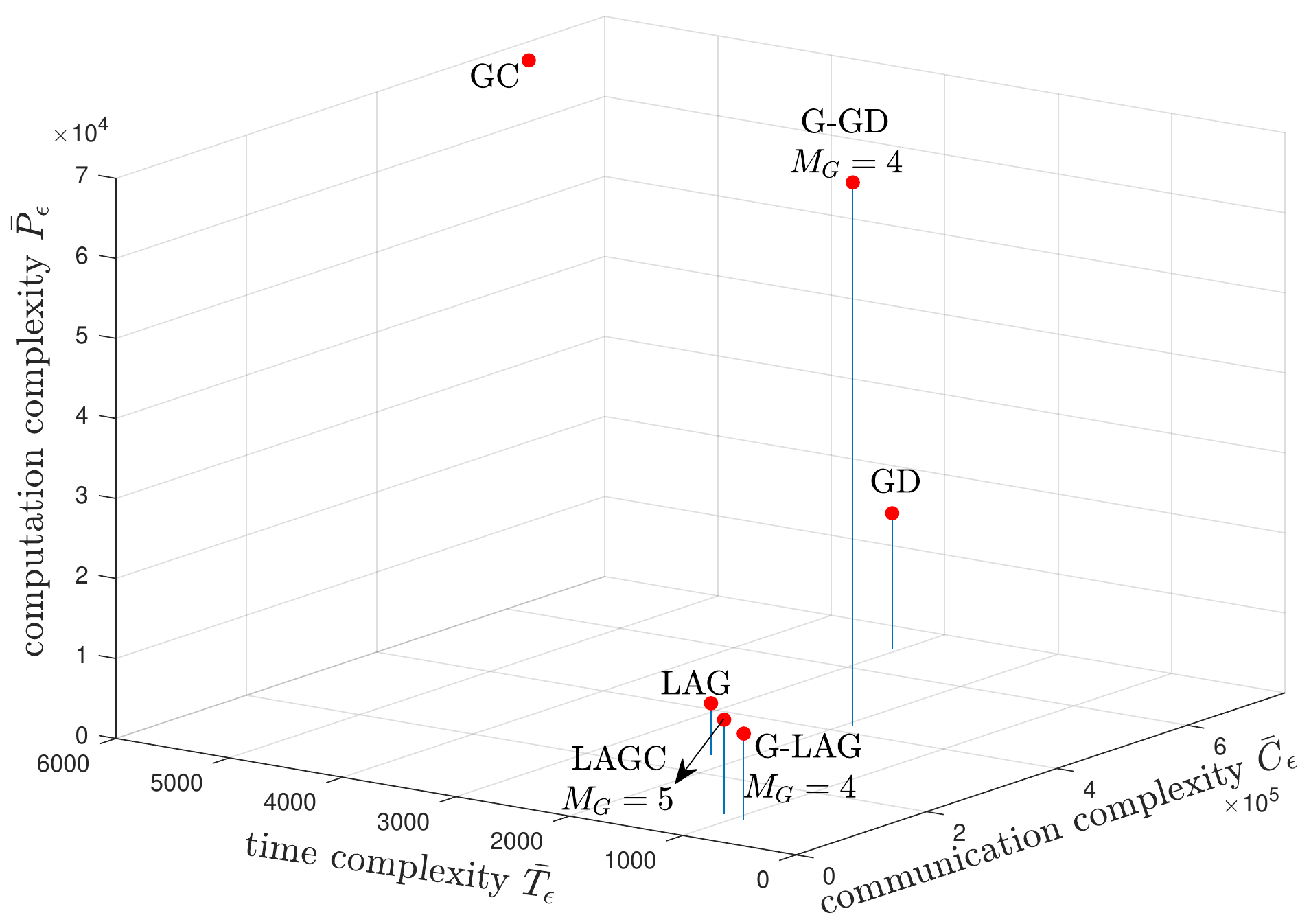}
\caption{Time, communication, and computation complexity measures for GD, GC, LAG and LAGC that guarantees the $\epsilon$-optimality gap with $\epsilon=10^{-8}$ under the exponential distribution for the workers' computing times. }
\label{fig:complexity_ex_3D}
\end{figure}

We now provide an illustration of the relative performance of the considered schemes. To this end, we consider a linear regression model $y=\pmb{\theta}^T \mathbf{x}$, where input $\mathbf{x}$ is an input vector obtained by pre-processing a vectorized $784\times 1$ handwritten digit image from the MNIST dataset. Specifically, the MNIST training dataset of 60,000 examples is divided into $S=20$ partitions of equal size, each including $3,000$ images for each digit $0,1,\dots,9$. Each partition $\mathcal{D}_s$, with $s=1,\dots,S$ is processed as in \cite{TGSY:18} such that the smoothness constant of the corresponding loss function is set to be equal $L_s=(1.3^{s-1}+1)^2$. The loss function $L_s(\pmb{\theta}) $ is given as $L_s(\pmb{\theta})=||\mathbf{X}_s \pmb{\theta} - \mathbf{y}_s||^2$, where $\mathbf{X}_s$ is the $3,000\times 784$ input matrix for the $s$th partition $\mathcal{D}_s$ (with each input corresponding to a row of $\mathbf{X}_s$) and the target vector $\mathbf{y}_s$ is given as $\mathbf{y}_s=\mathbf{X}_s \pmb{\theta}^*$ for a randomly generated ground-truth parameter vector $\pmb{\theta}^*$ with i.i.d. zero-mean unitary power Gaussian entries.  
We consider the most recent $D=10$ iterations to approximate the gradient norm square. Furthermore, for the computing times, the Pareto distribution has shape parameter $\alpha=1.1$ and scale parameter $0.005/\alpha$, and the exponential distribution has mean parameter $0.05$. We evaluate the complexity measures derived above ---  using the bounds \eqref{time:LAG}, \eqref{time:LAGC},  \eqref{compu:LAG}, \eqref{compu:LAGC} for LAG and LAGC --- for redundancy $r=4$ for GC and LAGC, as well as hyperparameters $\eta=0.05$, $\beta=1.1$, and $\xi=1$. We also set $F=17$ for GC and $F=M_G-r_G+1$ for G-GD and LAGC with $M_G$ set to different values. Time, communication and computation complexities of GD, GC, LAC, G-GD, and LAGC for different values of $M_G$ with Pareto distribution and exponential distribution are shown in Fig.~\ref{fig:complexity_pa_3D} and Fig.~\ref{fig:complexity_ex_3D}, respectively. As discussed, in both figures, when $M_G=1$ and $M_G=M$, LAGC coincides with LAG and GC, respectively, while, when for $M_G\leq r=4$, we have the (uncoded) G-LAG scheme in Remark \ref{GLAG}.

We first consider computing times with Pareto distribution. The high tail of the Pareto distribution entails a high probability that some workers are significantly slower than the rest. As seen in Fig.~\ref{fig:complexity_pa_3D}, in this case, both G-GD and GC have a lower time complexity than both GD and LAG thanks to their robustness to stragglers (recall Table~\ref{advan}): Although each active worker executes more computations, the reduced requirements on the number of workers that need to complete their computations offset the increased per-server computation load. However, this wall-clock time saving implies a trade-off with the computation complexity, which is increased. Furthermore, G-GD outperforms GC in all metrics, since more stragglers can be tolerated thanks to grouping for without increasing the computational redundancy. LAG has a larger wall-clock time complexity as compared to G-GD and GC, but it can significantly reduce both communication and computation complexities by selecting a reduced number of workers to be active. The proposed LAGC scheme is seen to be able to harness both the robustness to stragglers of GC and G-GD, which requires a larger $M_G$, and the reduced communication and computation complexity of LAG, which requires a smaller $M_G$. In fact, in line with the comparison between G-GD and GC, we observe that G-LAG --- the special case of LAGC that only with grouping and adaptive selection --- yields the best overall performance.

We now consider the performance under the exponential distribution for the workers' computing times. This distribution has a lower tail and hence the workers have comparable computing times with a higher probability than for the Pareto distribution. In this case, in stark contrast to Fig.~\ref{fig:complexity_pa_3D}, Fig.~\ref{fig:complexity_ex_3D} shows that GC does not improve the time complexity, since the cost resulting from computational redundancy does not offset the savings accrued thanks to the mitigation of stragglers. As compared to LAG, G-GD provides a reduction in wall-clock time due to its stronger ability to tolerate stragglers, implying a trade-off with the increasing computation complexity. LAG outperforms schemes based solely on grouping or coding in terms of communication and computation complexity. Finally, the proposed G-LAG outperforms all other schemes in terms of computation and time complexities while requiring a larger computation complexity than LAG.

 %Finally, G-LAG can provide additional savings in terms of time and communication complexities by selecting larger values of $M_G$ but at the cost of a larger computation complexity. For instance, with $M_G=4$, 

\begin{figure}
\centering
\subfigure{%
\includegraphics[width=0.5\textwidth]{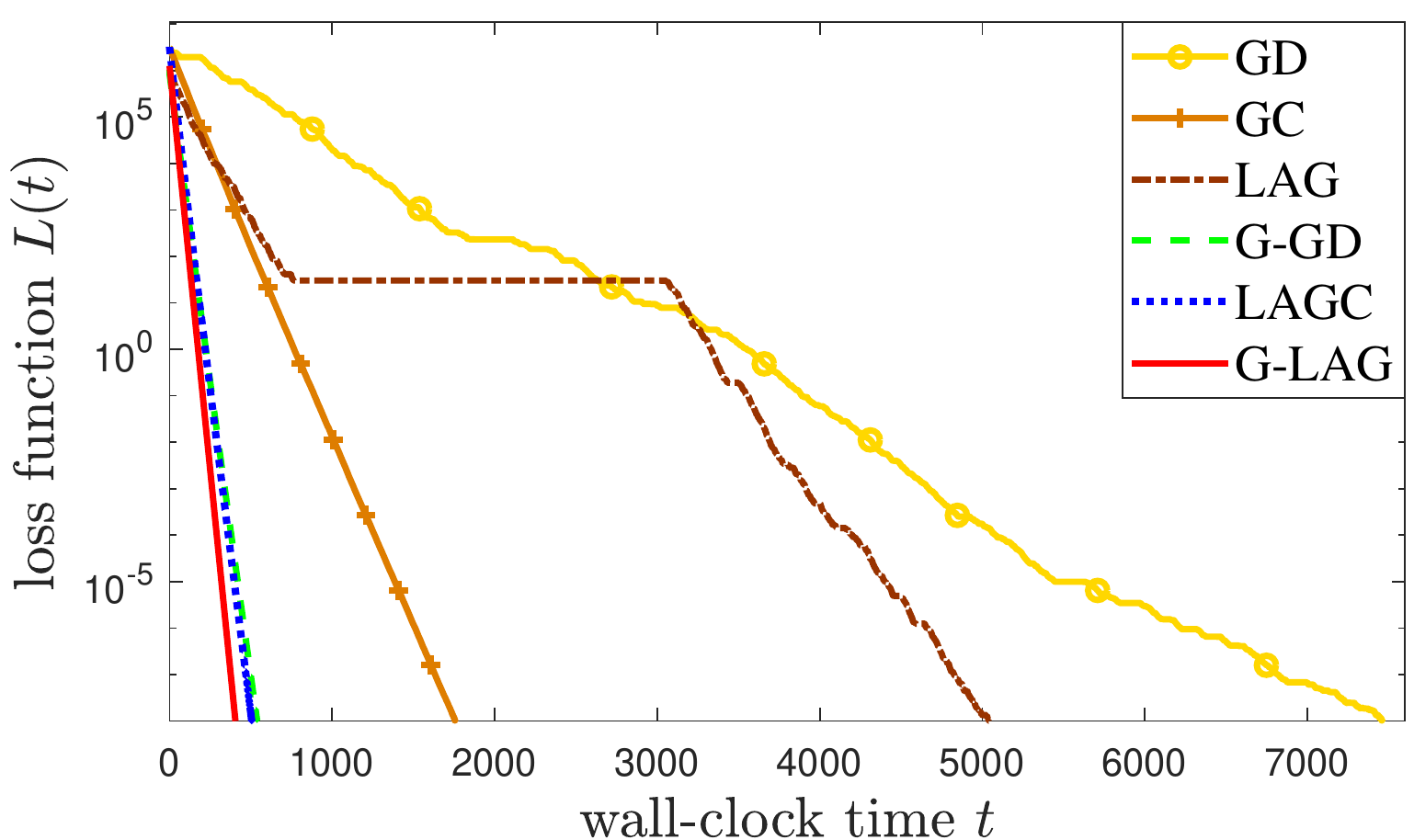}}%
\qquad
\subfigure{%
\includegraphics[width=0.49\textwidth]{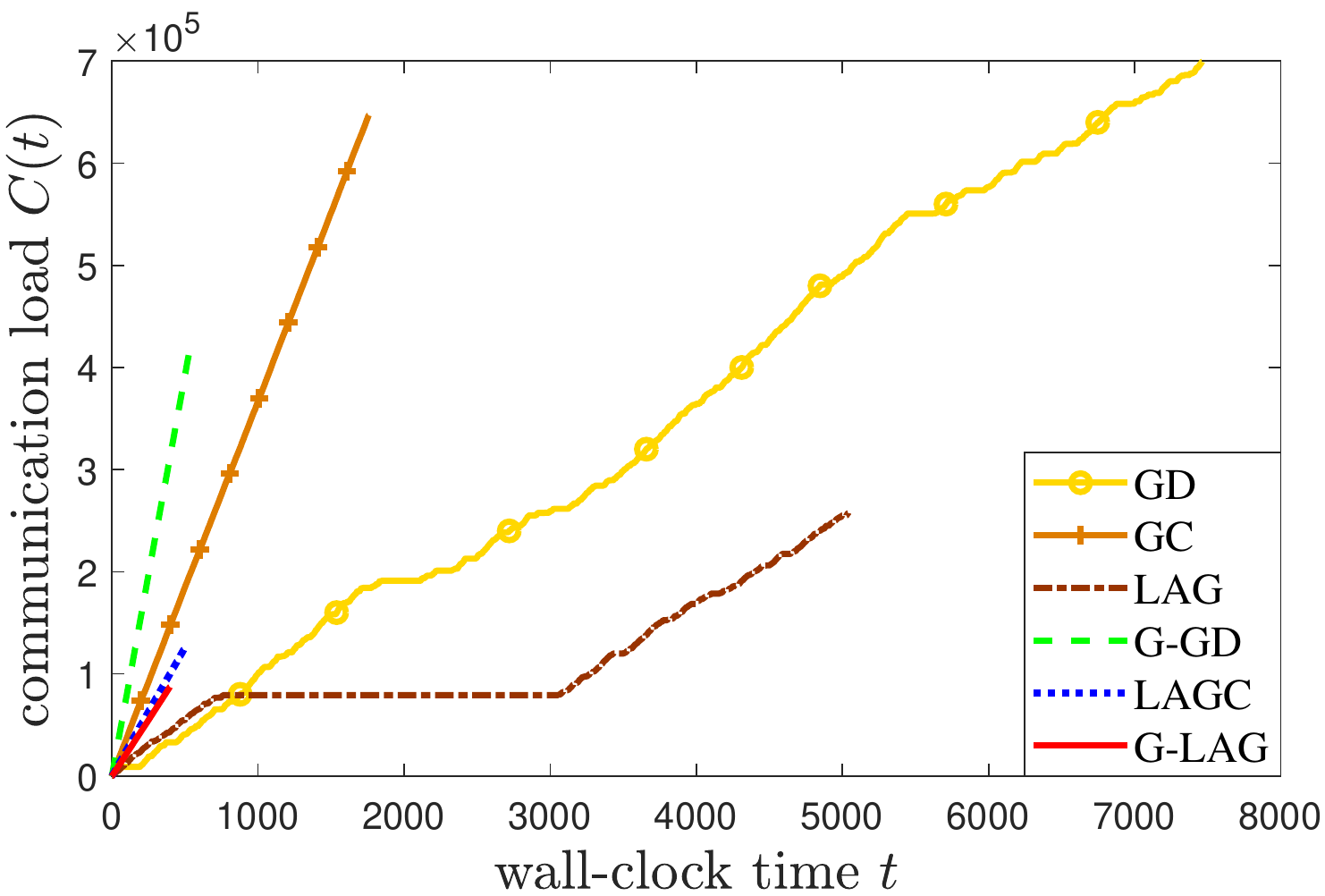}}%
\qquad
\subfigure{%
\includegraphics[width=0.49\textwidth]{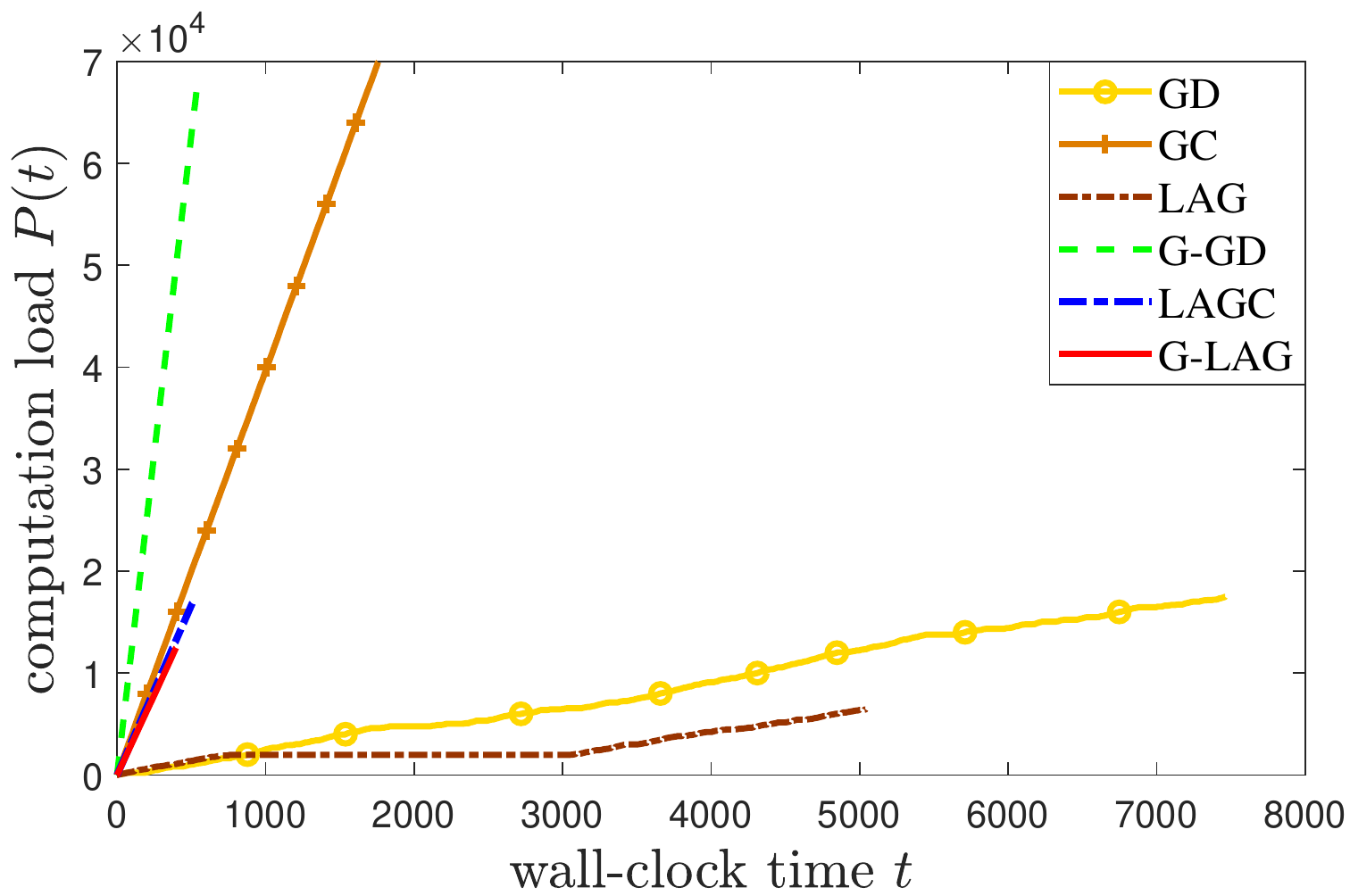}}%
\caption{Loss function $L(t)$, communication load $C(t)$, and computation load $P(t)$ versus wall-clock time $t$ with the Pareto distribution for the computing times.}
\label{fig:load_pa}
\end{figure}

	\begin{figure}%
\centering
\subfigure{%
\includegraphics[width=0.5\textwidth]{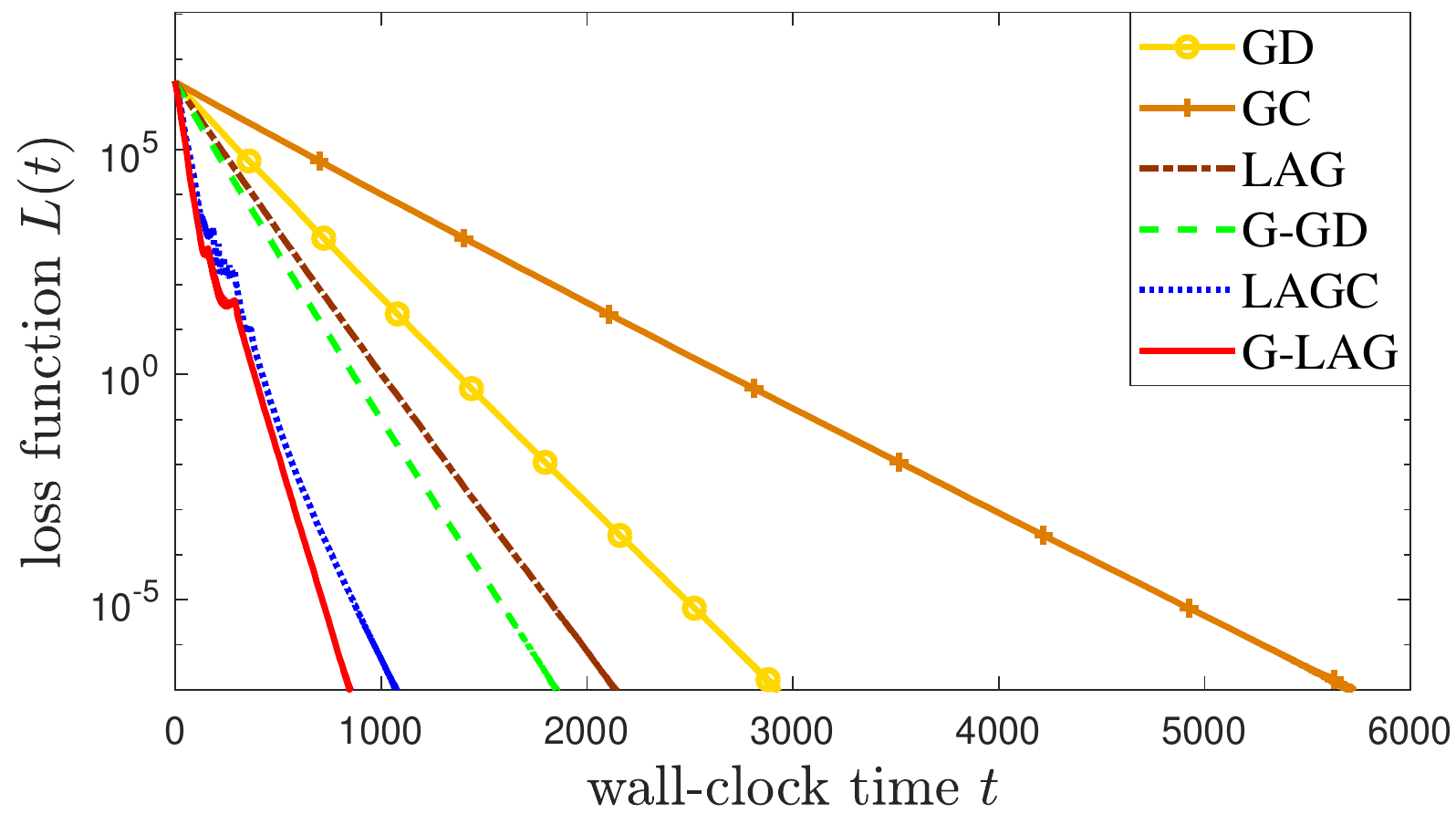}}%
\qquad
\subfigure{%
\includegraphics[width=0.5\textwidth]{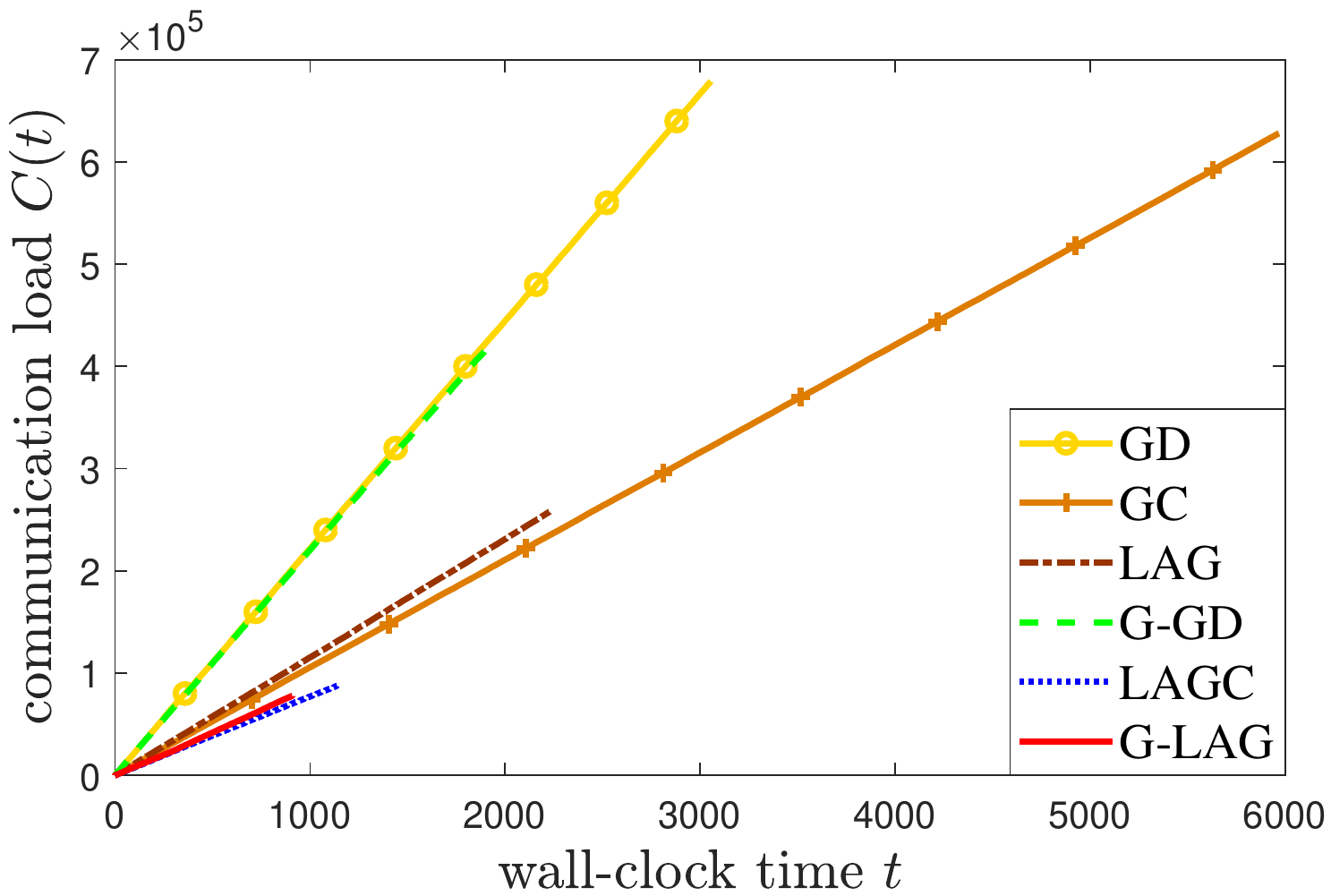}}%
\qquad
\subfigure{%
\includegraphics[width=0.5\textwidth]{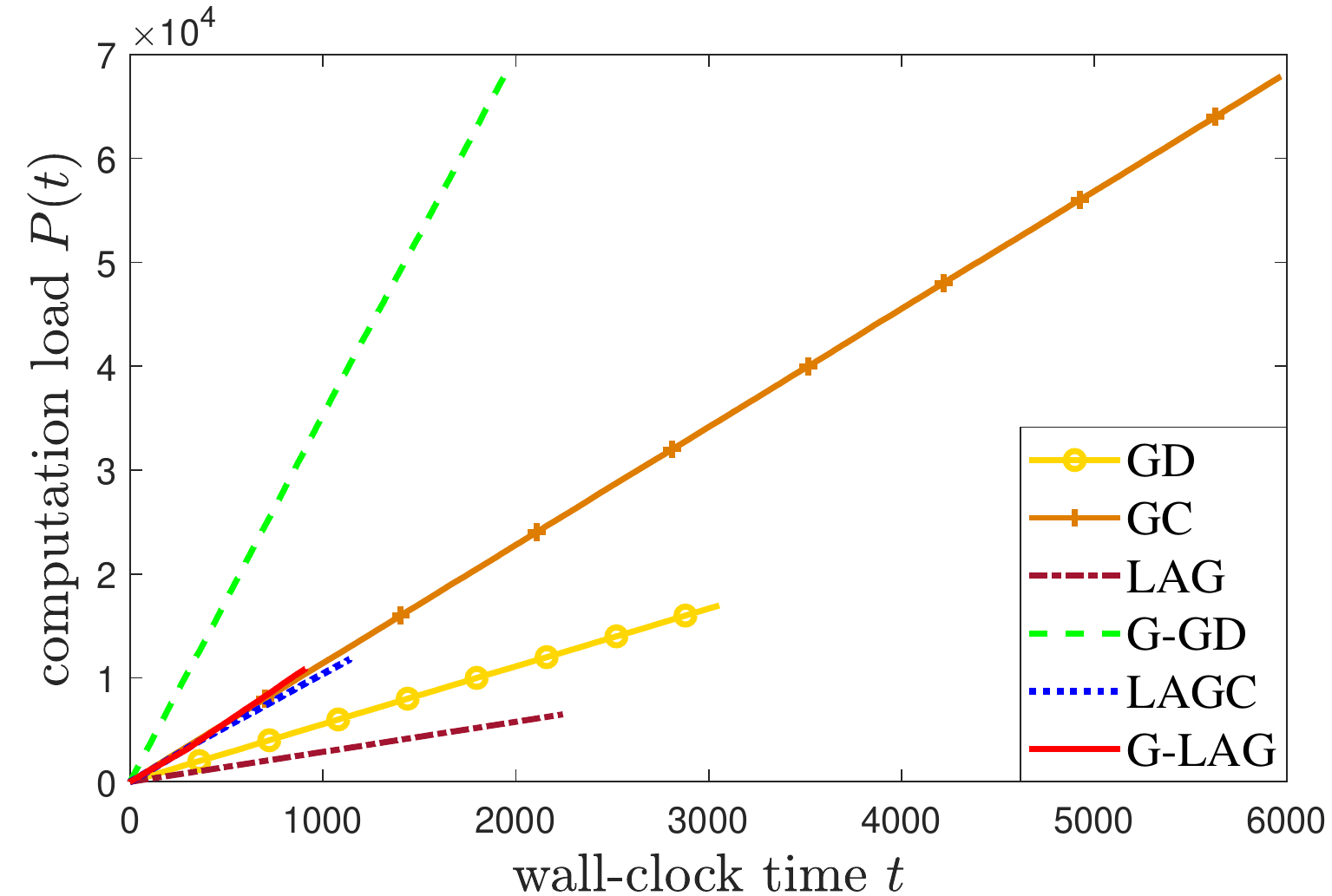}}%
\caption{Loss function $L(t)$, communication load $C(t)$, and computation load $P(t)$ versus wall-clock time $t$ with the exponential distribution for the computing times.}
\label{fig:load_ex}
\end{figure}

\section{Numerical Results}  \label{numerical}
In this section, we present numerical examples in order to illustrate the loss function $L(t)$ in \eqref{def:loss}, communication load $C(t)$ in \eqref{load:comm}, and computation load $P(t)$ in \eqref{load:comp} as function of wall-clock time $t$ for the considered training strategies. We adopt the same linear regression set-up described in Section~\ref{illustra} with Pareto and exponential distributions for the random computing times of the workers. In Fig~\ref{fig:load_pa} and \ref{fig:load_ex}, we plot the mentioned metrics averaged over 100 random realization of the computing times for GD, GC, LAG, G-GD, LAGC with $M_G=5>r$, and G-LAG with $M_G=r$. We note that each curve terminates at the time when the $\epsilon$-optimality is achieved.
%Furthermore, at any rumtime $t$, the faster scheme that stops first completes more iterations since all the schemes have the same order of iteration complexity.
%We consider a synthetic dataset with increasing smoothness constants, given as $L_m=(1.3^{m-1}+1)^2$, with $m\in[M]$. For each group, 100 data samples with $\xv_n\in \R^{100}$ is generated from a standard Gaussian distribution. 

%hence the average run-time per-iteration is scaling as the 
%
%hence at any wall-clock time $t$, they have completed different number of iterations if the wall

%\begin{figure}[t!] 
  %\centering
%\includegraphics[width=0.43\columnwidth]{error_pa.pdf}
%\caption{Loss function $L(t)$ vs wall-clock time $t$ with the Pareto distribution.}
%\label{fig:error_pa}
%\end{figure}
%

Confirming the conclusion from the analysis in Section~\ref{illustra}, under Pareto computing times, the loss functions of GC, G-GD, LAGC, and G-LAG decrease significantly faster than GD and LAG thanks to their robustness to stragglers, as shown in Fig.~\ref{fig:load_pa}, with G-LAG yielding the steepest descent. LAG is seen to be effective in reducing the communication and computation loads per unit of time. However, G-LAG yields a small overall communication complexity, at the cost of a large computation due to the smaller time for iteration afforded by its robustness to stragglers.

  %and LAGC and G-LAG converge faster than GC by selecting the active groups, the same as in Fig~\ref{fig:complexity_pa_3D} for time complexity. In contrast, GD is extremely slow due to the fact that all the workers, including the significantly slow ones, are active at each iteration. LAG is better than GD by selecting only a subset of active workers. On the other hand, as illustrated in Fig~\ref{fig:load_pa}(a) and (b), GC, LAGC, and G-LAG end up with higher communication and computation loads. This is because, at any run-time $t$, they have completed more iterations than GD and LAG since all the schemes have the same order of iteration complexity but they are faster. Finally, computation loads have the similar behavior because of the computational redundancy of GC, LAGC, and G-LAG. It should be noted that the end points of the communication and computation loads are consistent with the points in Fig~\ref{fig:complexity_pa_3D}.

We now turn to considering exponential computing times. In this regard, Fig~\ref{fig:load_ex} verifies the conclusion based on the analysis in Section~\ref{illustra} that coding is not advantageous in terms of any performance metric with respect to GC. In contrast, G-GD can be significantly more time efficient due to its stronger robustness to stragglers. By using adaptive selection, the communication loads of LAG, LAGC and G-LAG increase with similar rates as a function of $t$, but G-LAG has a smaller communication complexity due to the smaller time complexity. Finally, due to computational redundancy, G-LAG, GC, and LAGC have higher computation loads than the other schemes.

\section{Conclusions} \label{conclusion}
%summarize the main ideas with reference to table I.
In this work, we explored the trade-off among wall-clock time, communication, and computation requirements for gradient-based distributed learning by leveraging coding, grouping, and adaptive selection. As summarized in Table~\ref{advan}, both coding and grouping provide robustness to stragglers, while adaptive selection is beneficial to reduce communication and computation loads. We proposed two novel strategies that aim at integrating the benefits of both types of approaches. Through analysis and numerical results, we have concluded that, when the distribution of the computing times of the workers has a low tail, the advantage of straggler mitigation via coding does not compensate for the increased computation load even in terms of wall-clock run-time. In contrast, for both high- and low-tail distributions of the computing times, the proposed G-LAG was seen to strike a desirable balance in terms of wall-clock time and communication overhead, with only a limited increase in computation cost.

This work leaves open a number of research directions. First, it would be interesting to combine stochastic gradient coding \cite{NRDI:18,CPE:17,BWR:19} with grouping and adaptive selection. Second, the potential advantages of the techniques considered here should be reconsidered for asynchronous implementations, where any server can compute the gradient and send an update to the PS without waiting for the other servers \cite{DGGDN:17}. Lastly, a related issue would be to introduce data privacy requirements \cite{BVBMMPRS:17}.

\begin{appendices}
\section{Proof of bounds~\eqref{time:LAG} and~\eqref{time:LAGC}} \label{sec:proof}

Define as $F(x)$ the Cumulative Distribution Function (CDF) of each variable $T_i$. From \cite[Lemma 2]{HAD:97}, we have the equality 
%\begin{align}
%\mathrm{E}[T_{a+1:b}]- \mathrm{E}[T_{a:b-1}]=\binom{b-1}{a}\int^{+\infty}_{-\infty} F^a(x)[1-F(x)]^{b-a}dx.
%\end{align}
%Let $a=b-1$, i.e., consider the largest expected order statistics, then the above equality becomes
\begin{align} \label{appen:1}
\bar{T}_a-\bar{T}_{a-1}=\int^{+\infty}_{-\infty} F^{a-1}(x)[1-F(x)]dx.
\end{align}
From \eqref{appen:1}, 
%\begin{align} \label{appen:2}
%\mathrm{E}[T_{b+i:b+i}]-\mathrm{E}[T_{b-1+i:b-1+i}]=\int^{+\infty}_{-\infty} F^{b-1+i}(x)[1-F(x)]dx.
%\end{align}
since $0\leq F(x)\leq 1$, the function $\bar{T}_a$ has decreasing increments in $a$, and hence it is discrete concave \cite{M:03}. It follows that for any integers $a_i\leq b$ with $\sum_{i=1}^{K}a_i=S_a$, we have Jensen's inequality 
\begin{align} \label{jensen}
\bar{T}_{\frac{S_a}{K}}\geq \frac{1}{K}\sum_{i=1}^{K}\bar{T}_{a_i},
\end{align}
as long as $S_a/K$ is an integer.
%\begin{align}\label{appen:3}
%\mathrm{E}[T_{a:a}]+\mathrm{E}[T_{a':a'}]\geq \mathrm{E}[T_{a-1:a-1}]+\mathrm{E}[T_{a'+1:a'+1}],
%\end{align}
%$2\mathrm{E}[T_{b:b}]\geq \mathrm{E}[T_{b-1:b-1}]+\mathrm{E}[T_{b+1:b+1}]$, 
%Applying this inequality recursively, we can then prove the inequality 
%\begin{align} \label{prof:1}
%K\mathrm{E}[T_{S_a/K:S_a/K}] \geq \sum_{i=1}^{K} \mathrm{E}[T_{a_i:a_i}]
%\end{align}
%for any integer $a_i\leq b$ with $\sum_{i=1}^{K}a_i=S_a$. 
In order to apply \eqref{jensen}, we define $a_i=|\mathcal{M}_U^i|$ and $K=|\mathcal{I}_{\epsilon}^{LAG}|$. We then have the desired inequality in \eqref{time:LAG}. Bound \eqref{time:LAGC} follows from the same argument.

\end{appendices}

 \section*{Acknowledgements}
Jingjing Zhang and Osvaldo Simeone have received funding from the European Research Council (ERC) under the European Union's Horizon 2020 Research and Innovation Programme (Grant Agreement No. 725731). 

\bibliographystyle{IEEEtran}
\bibliography{IEEEabrv,final_refs}

\end{document}

%% file: defs.tex
\newcommand{\bit}{\begin{itemize}}
\newcommand{\eit}{\end{itemize}}

\newcommand{\bc}{\begin{center}}
\newcommand{\ec}{\end{center}}

\newcommand{\ba}{\begin{array}}
\newcommand{\ea}{\end{array}}

\newcommand{\beq}{\begin{equation}}
\newcommand{\eeq}{\end{equation}}

\newcommand{\beqn}{\begin{equation*}}
\newcommand{\eeqn}{\end{equation*}}

\newcommand{\bean}{\begin{eqnarray*}}
\newcommand{\eean}{\end{eqnarray*}}
\newcommand{\bea}{\begin{eqnarray}}
\newcommand{\eea}{\end{eqnarray}}

\def\R{\mathbb{R}}

\def\F{\mathbb{F}}
\def\fv{\boldsymbol{f}}
\def\gv{\boldsymbol{g}}

\def\xv{\boldsymbol{x}}

\def\zv{\boldsymbol{z}}
\newtheorem{remark}{Remark}